\documentclass[final]{IEEEtran}
\usepackage{mathptmx} 
\usepackage[sort,nocompress]{cite}
\usepackage[keeplastbox]{flushend}
\usepackage{amsmath,amssymb,amsfonts}
\usepackage[noend]{algorithmic}
\usepackage[]{algorithm}
\usepackage{graphicx}
\usepackage{textcomp}
\usepackage{xcolor}
\usepackage{fancyhdr}
\usepackage[normalem]{ulem}
\usepackage{xspace}
\usepackage{tikz}
\usepackage{hyperref}
\usepackage{enumitem}
\usepackage{pifont}

\def\BibTeX{{\rm B\kern-.05em{\sc i\kern-.025em b}\kern-.08em
    T\kern-.1667em\lower.7ex\hbox{E}\kern-.125emX}}
    
\newcommand*{\circled}[1]{\lower.7ex\hbox{\tikz\draw (0pt, 0pt)%
    circle (.5em) node {\makebox[1em][c]{\small #1}};}}
\newcommand{\yt}[1]{\textcolor{red}{#1}}

\newcommand{\design}{{\textsc{NDSearch}}\xspace}
\newcommand{\ssd}{{\textsc{SearSSD}}\xspace}

\title{\design: Accelerating Graph-Traversal-Based Approximate Nearest Neighbor Search through Near Data Processing} 

\author{
\IEEEauthorblockN{Yitu Wang\IEEEauthorrefmark{2}, Shiyu Li\IEEEauthorrefmark{2}, Qilin Zheng\IEEEauthorrefmark{2}, Linghao Song\IEEEauthorrefmark{3}, \\ Zongwang Li\IEEEauthorrefmark{4}, Andrew Chang\IEEEauthorrefmark{4}, Hai ``Helen'' Li\IEEEauthorrefmark{2}, Yiran Chen\IEEEauthorrefmark{2}}

\IEEEauthorblockA{\IEEEauthorrefmark{2}Duke University, Durham, North Carolina, USA}   

\IEEEauthorblockA{\IEEEauthorrefmark{3}University of California, Los Angeles, California, USA}

\IEEEauthorblockA{\IEEEauthorrefmark{4}Samsung Semiconductor, Inc., San Jose, California, USA }

\IEEEauthorblockA{
\url{yitu.wang@duke.edu} \hspace{1em}
\url{shiyu.li@duke.edu}
\hspace{1em}
\url{qilin.zheng@duke.edu}
\hspace{1em}
\url{linghaosong@cs.ucla.edu}
\\
\url{zongwang.li@samsung.com}
\hspace{1em}
\url{andrew.c1@samsung.com}
\hspace{1em}
\url{hai.li@duke.edu}
\hspace{1em}
\url{yiran.chen@duke.edu}}
\vspace{-2em}}

\begin{document}
\maketitle
\thispagestyle{plain}
\pagestyle{plain}


\begin{abstract}
Approximate nearest neighbor search~(ANNS) is a key retrieval technique for vector database and many data center applications, such as person re-identification and recommendation systems. It is also fundamental to retrieval augmented generation (RAG) for large language models (LLM) now.
Among all the ANNS algorithms, graph-traversal-based ANNS achieves
the highest recall rate.
However, as the size of dataset increases, the graph may require hundreds of gigabytes of memory, exceeding the main memory capacity of a single workstation node.
Although we can do partitioning and use solid-state drive (SSD) as the backing storage, the limited SSD I/O bandwidth severely degrades the performance of the system.
To address this challenge, we present \design, a hardware-software co-designed near-data processing~(NDP) solution for ANNS processing.
\design consists of a novel in-storage computing architecture, namely, \ssd, that supports the ANNS kernels and leverages logic unit~(LUN)-level parallelism inside the NAND flash chips.
\design also includes a processing model that is customized for NDP and cooperates with  \ssd.
The processing model enables us to apply a two-level scheduling to improve the data locality and exploit the internal bandwidth in \design, and a speculative searching mechanism to further accelerate the ANNS workload.
Our results show that \design improves the throughput by up to 31.7$\times$, 14.6$\times$, 7.4$\times$ 2.9$\times$ over CPU, GPU,  a state-of-the-art SmartSSD-only design, and DeepStore, respectively. \design also achieves two orders-of-magnitude higher energy efficiency than CPU and GPU.
\end{abstract}

\section{Introduction}
\label{sec:intro}
Approximate nearest neighbor search (ANNS) is the fundamental technique of the similarity search in the vector database~\cite{vectordatabaseintro,milvus} and has been applied to a wide range of significant application domains~\cite{annsappdomain}, including pattern recognition~\cite{patternrecog1,patternrecog2}, machine learning~\cite{ml1,ml2, wen2019memristor,wang2020reboc,chen2018regan}, information retrieval~\cite{inforretr1,inforetr2,zhu2016deep,barman2019graph}, data mining~\cite{datamining1,datamining2,hnswamazon} and recommendation system~\cite{recsys1,recsys2,wang2021rerec,wang2023ems,li2024ndrec}. Currently, ANNS has also been applied to retrieval augmented generation (RAG)~\cite{rag}, which provide the relevant information for the enhanced context of the Large Language Model (LLM) ~\cite{llm}.
In these applications, a query is usually processed by two stages~\cite{two-stage-recsys,two-stage-reid}: retrieve/recall stage and rank/identification stage. During the first stage, a fixed number of neighbors are retrieved from the database. Then, during the second stage, the retrieved data of the given query is further delicately processed by the specialized models.
Instead of getting the exact nearest neighbors, ANNS boosts the search speed by limiting the range of candidates and sacrificing some recall rate.
Since the modern vector database and applications deal with large-scale data, boosting the performance of ANNS is critical.

Among existing methods~\cite{arya1998optimal, andoni2006near, fu2019fast} of ANNS, graph traversal-based methods such as hierarchical navigable small world graphs (HNSW)~\cite{malkov2018efficient} and DiskANN~\cite{subramanya2019diskann} are the most popular ones for the optimal recall-throughput tradeoff that they achieved~\cite{cvpr20_tutorial_image_retrieval}. 
These methods construct a graph based on the distance between the feature vector of each vertex in the dataset.
The search is then performed by traversing the neighboring vertices of the visited ones until the predefined condition is met.
The graph traversal-based ANNS  consists of three steps: \textit{graph traversal}, \textit{distance computation}, and \textit{bitonic sorting}. 
Graph traversal searches for the potential nearest neighbors of the given queries; 
distance computation calculates the Euclidean/angular distance between the visited vertices and the queries; 
and bitonic sorting~\cite{chen2015energy} generates the top-k candidates of each query in a batch.
\begin{figure}[t]
	\begin{center}	
        \vspace{0em}
		\includegraphics[width=\columnwidth]{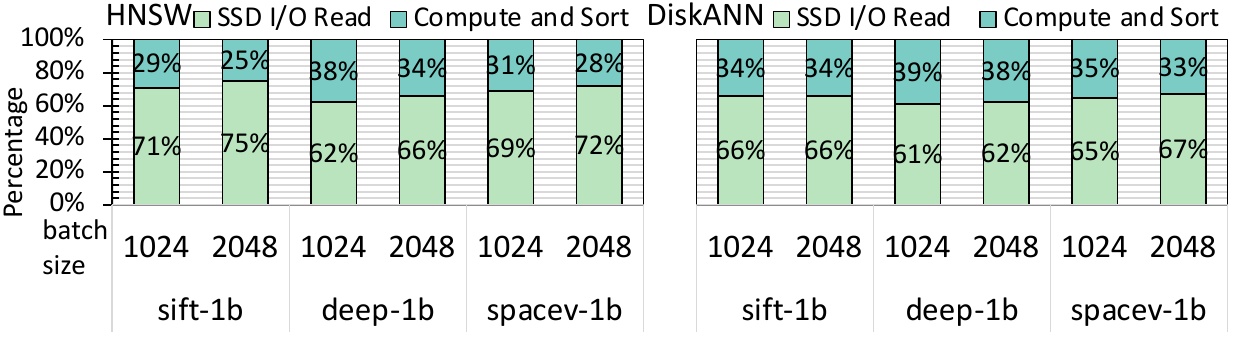}
		\vspace{-1.5em}
		\caption{Execution time breakdown of HNSW and DiskANN running on two Intel Xeon Gold 6245 CPUs.}
        \vspace{-2em}
		\label{AccessSSD}
	\end{center}	

\end{figure} 
In large-scale real-world applications, a graph can contain up to billions of vertices~\cite{bigann}.
Considering the feature vector and the adjacency information associated with each vertex, the workload may consume a significant amount of memory.
For example, in HNSW, the memory consumption per vertex ranges from 60 bytes to 450 bytes~\cite{malkov2018efficient}.
Given the large number of vertices, the required memory of ANNS could reach hundreds of gigabytes (GBs) or even several terabytes (TBs), which exceeds the total capacity of the main memory on a single node in the workstation~\cite{diskann}. 

\begin{figure}[t]
	\begin{center}	
        \vspace{0em}
		\includegraphics[width=\columnwidth]{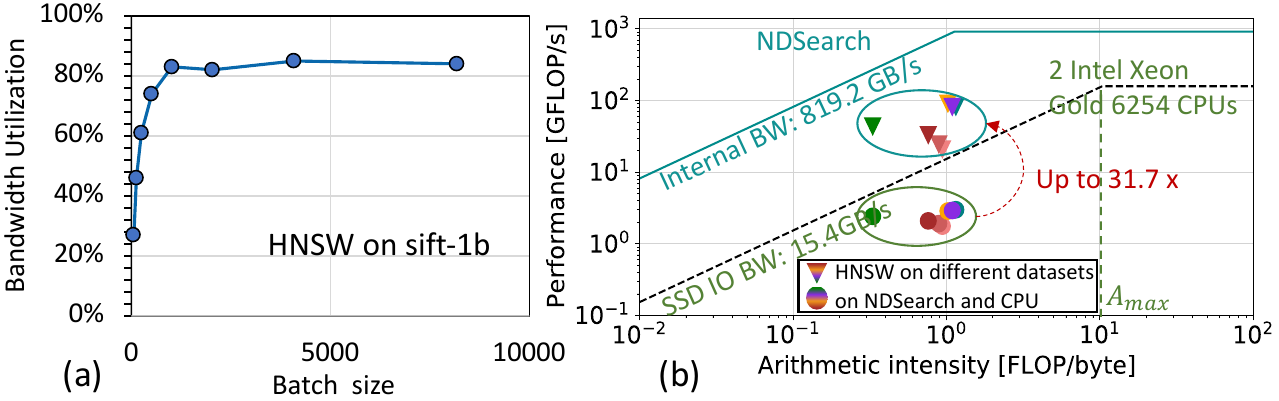}
		\vspace{-1.5em}
		\caption{(a) PCIe bandwidth saturates as the batch size increases; (b) Roofline lifting effect and the ANNS workloads speedup enabled by \design (all the page buffers can be read simultaneously). }
		\label{Roofline}
	\end{center}	
	\vspace{-2.5em}
\end{figure}

Existing solutions generally adopt three approaches to fulfill the memory capacity demand: (\romannumeral1) sharding the dataset, storing the shards in the disk, loading a limited number of shards into memory, and only routing the query to these shards to perform the search locally~\cite{hnswlib}; (\romannumeral2) developing programs with SSD-based indices to fetch the feature vectors from SSD into the main memory at runtime~\cite{diskann}; (\romannumeral3) loading the whole dataset from the SSD into multiple nodes' main memory at the cost of expensive machines and high power consumption, e.g., accelerating ANNS with 8 A100 GPUs~\cite{GPUAcc}.
Although these designs can support graph-traversal-based ANNS on very large datasets, their performance is greatly limited by the SSD I/O read (the time taken by data transfer via PCIe) in terms of the end-to-end performance, as shown in Fig.\ref{AccessSSD}. The SSD I/O read accounts for up to 75\% of the total latency. From Fig.\ref{Roofline}(a), we can see that the  utilization of SSD I/O bandwidth saturates to 83\%, after the batch size increases to 1024. Combined with Fig.~\ref{AccessSSD}, the saturated SSD I/O bandwidth illustrates that the SSD I/O read latency comes from the limited I/O bandwidth. In the current computer systems, CPU and SSD are usually connected via PCIe links, e.g., PCIe 3.0 $\times$ 16 whose peak bandwidth is about 15.4 GB/s. Fig.~\ref{Roofline}(b) further shows that the ANNS workloads are located in the SSD I/O bandwidth-constrained region. We conclude that the SSD I/O bandwidth limits the performance of ANNS. To fundamentally address the issue, we envision that a promising solution is to \textit{directly process the large-scale ANNS workload within the storage}. 


However, it is not straightforward to implement graph traversal-based ANNS within the storage devices.
From the hardware side, current in-storage accelerators like DeepStore~\cite{mailthody2019deepstore} are unable to fully utilize the internal SSD bandwidth for graph traversal-based ANNS because ANNS's  irregular and sparse data access pattern require more fine-grained level parallelism.
From the software side, current processing models of graph traversal~\cite{sundaram2015graphmat,yan2019alleviating} lacks the applicability to the in-storage architectures, especially the customized dataflow and scheduling.
To tackle the above challenges, we propose \design, a hardware-software co-deisgned NDP solution for ANNS based on SmartSSD \cite{lee2020smartssd}.
The operations of the graph traversal and distance computation kernels are offloaded to the SSD.
We studied the data access pattern of graph traversal on SSDs and found that the fine-grained and in-place acceleration at the logic unit (LUN) level is the most promising solution for processing the irregular and scattered accesses to feature vectors in graph-traversal-based ANNS. Hence, we develop LUN-level accelerators on flash chips with the modified multi-LUN operations.
Meanwhile, the high parallel bitonic sorting  is offloaded onto the FPGA like ~\cite{salamat2021nascent} to allow more power and area budget.
To fully exploit the proposed architecture, \design also includes a novel processing model designed for ANNS in NDP scenarios.
The processing model is based on GraphMat~\cite{sundaram2015graphmat} and incorporates a two-level scheduling scheme, static and dynamic scheduling. 
Static scheduling reorders and maps the vertices to the physical location with the goal of enabling multi-plane operations and maximizing the \underline{spatial locality}.
Based on a new graph format, LUNCSR, dynamic scheduling aims at maximizing the \underline{temporal locality} of the data by allocating the queries based on the location of the required data.
A speculative searching mechanism is integrated into the dynamic scheduling to prefetch the vertices.

Our contribution can be summarized as follows.
   \begin{itemize}[leftmargin=*]
   \item We conduct a characterization study on graph-traversal-based ANNS and identify the opportunities of offloading the graph traversal and distance computation kernels to SSDs to address the issues of large memory consumption and limited SSD I/O bandwidth.
    \item From the perspective of hardware, we propose an NDP architecture design and a corresponding processing model to support graph-traversal-based ANNS.
    \item From the perspective of software, we develop a two-level scheduling scheme to both reorder the vertices in the SSD and dynamically allocate and prefetch queries during the search based on a new graph format, namely, LUNCSR.
\end{itemize}
The experimental results show up to 14.6$\times$ and 2.9$\times$ speedup over GPU and DeepStore, respectively. \design also achieves up to 30.06 $\times$ and 3.48 $\times$ higher energy efficiency than a state-of-art SmartSSD design~\cite{smartSSDKNN} and DeepStore~\cite{mailthody2019deepstore}.

\section{Background}
\label{sec:back}
\subsection{Graph-traversal-based ANNS}
\label{ANNSbackground}
ANNS is an approximation to the original nearest neighbor search (NNS) algorithm~\cite{fukunaga1975branch}, which aims to find top-K elements to a given query with minimized distances by scanning the whole dataset. 
Due to the high search cost of the brute-force search in NNS, ANNS is selected as an alternative and faster approach in scenarios where a lower recall rate is acceptable. 
Recently, graph-traversal-based ANNS such as HNSW~\cite{malkov2018efficient} and DiskANN~\cite{subramanya2019diskann} become popular due to their superior performance in the real life applications. Although the details of these two algorithms are different, the common basics can be generally summarized as follows.

\begin{figure}[t]
	\begin{center}	
         \vspace{0em}
		\includegraphics[width=0.8\columnwidth]{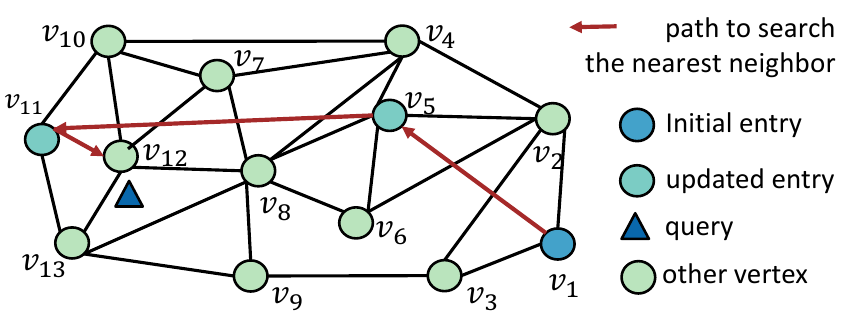}
		\vspace{-0.5em}
		\caption{The search phase of graph-traversal-based ANNS. }
		\label{HNSW}
	\end{center}	
	\vspace{-2.5em}
\end{figure} 
Graph-traversal-based ANNS consists of two phases, construction and search.
In the construction phase, a graph is constructed based on raw feature vectors in the dataset. Firstly, a distance function is defined to characterize the similarity between two vectors. Then, given a distance threshold and the maximum number of neighbors that one vector can have, incoming vectors are consecutively inserted in random entries and bidirectionally connected to their neighbors. 
Finally, vectors are built as vertices in the constructed graph. Each vertex consists of its feature vector and adjacency information (i.e., which neighbors the vertex is connected to and the distances between the vertex and the neighbors).
In the search phase, for a given query, a random entry vertex is firstly selected and put into the candidate list. Then the nearest vertex $C$ to the query is selected as the updated entry vertex from all the candidates and removed from the candidate list. If the distance between $C$ and the query meets a pre-defined condition or is far greater than a marked value in the result list, where the distances between the visited vertices and the query are stored, then the searching is terminated. Otherwise, add the neighbors (which are never visited) of $C$ to the candidate list. After traversing all the vertices in the candidate list and repeating the process above, the final top-K vertices from the results list are selected and sorted according to the distances in ascending order. Fig.~\ref{HNSW} shows a na\"ive example of searching the approximate nearest neighbor of the query. Our work mainly focuses on accelerating the search phase of ANNS, which is directly applied to various applications.





\subsection{SSD Preliminary}
\subsubsection{Internal organization}
Modern SSD consists of 2--4 embedded cores as the SSD controller, a few GBs of DRAM, and 16--32 channels of flash chips as the storage~\cite{eshghi2013ssd}.
The NAND flash storage elements are hierarchically organized at multiple levels - channels, chips, logic units (LUN), planes, blocks, and pages. 
Each channel contains a flash controller and 4--8 flash chips, each of which includes 2--8 planes. 
Each plane is made of a group of blocks and a page buffer.
Each block has multiple pages, whose size could be 2/4/8/16 KB. 
One or more planes are organized as a LUN~\cite{LUNorgnization,LUNorgnization2}, which is the minimal unit that can independently execute commands.
Multi-LUN~\cite{semiconductor2006open,LUNorgnization2} operations are supported in the flash chip to improve the in-chip parallelism.
Multi-plane operations can be supported to maximize the operation parallelism in a LUN.
The address of NAND flash consists of two parts: the column address and the row address~\cite{rowaddress}. 
The column address is used to access bytes or words within a page, while the row address is used to address pages, blocks, and LUNs.

\subsubsection{Data refreshing and address translation }
\label{sec:datarefreshing}
The flash transaction layer (FTL) runs on the embedded cores to process data refreshing, address translation, garbage collection and etc~\cite{chung2009survey}. Although the search phase of ANNS on \design does not induce graph updates and only operates in a read-only mode, the NAND flash requires data refreshing and data correction due to retention and read disturbance issues, resulting in the flash addresses being changed. We employ the block-level FTL refreshing mechanism and integrate the logical-to-physical address translation in the design of LUNCSR (see Section~\ref{LUNCSR} for the details). We use the specialized hardware component to infer the final physical address with the vertex index to eliminate the overhead of executing FTL address translation on the embedded cores. Compared to DeepStore~\cite{mailthody2019deepstore} which just considers the change of the starting address of the  database, our design is more realistic in terms of data refreshing and address translation.


\section{Motivation}
\label{motivation}
\begin{figure}[t]
     \centering
     \vspace{0em}
     \includegraphics[width=0.9\columnwidth]{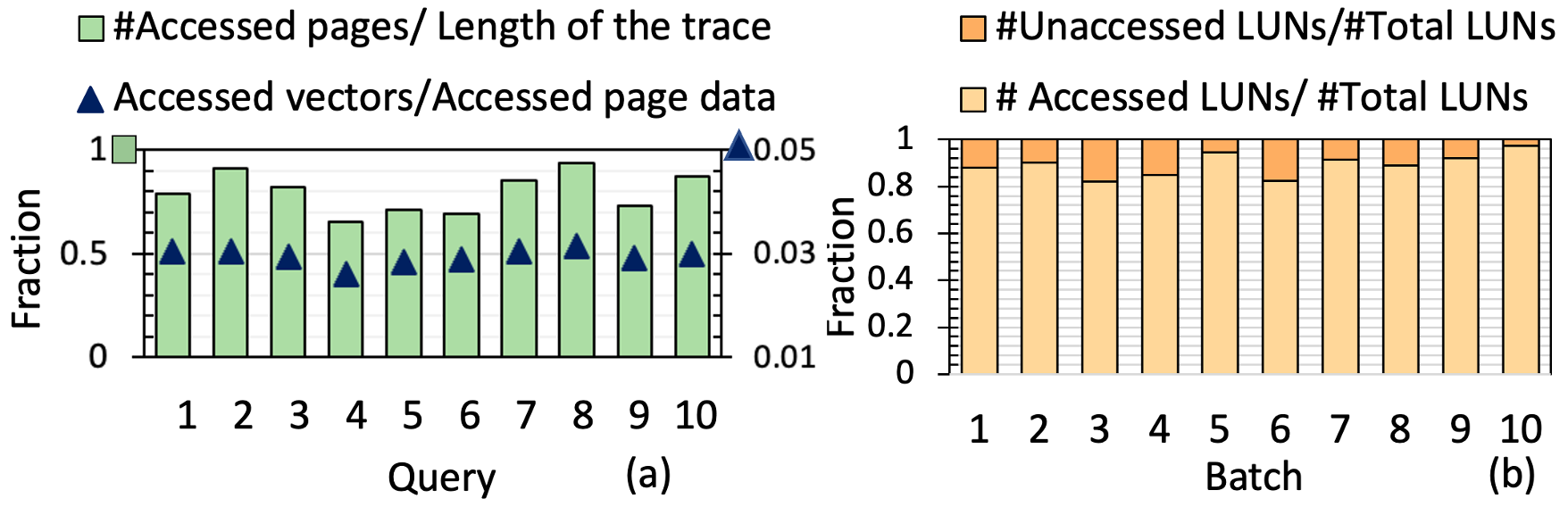}
     				\vspace{-1em}
     \caption{Page and LUN access pattern of the search phase.}
     \label{accesspattern}
     \vspace{-1em}
\end{figure}
\begin{figure*}[t]
	\begin{center}	
	    \vspace{0pt}
		\includegraphics[width=0.95\textwidth]{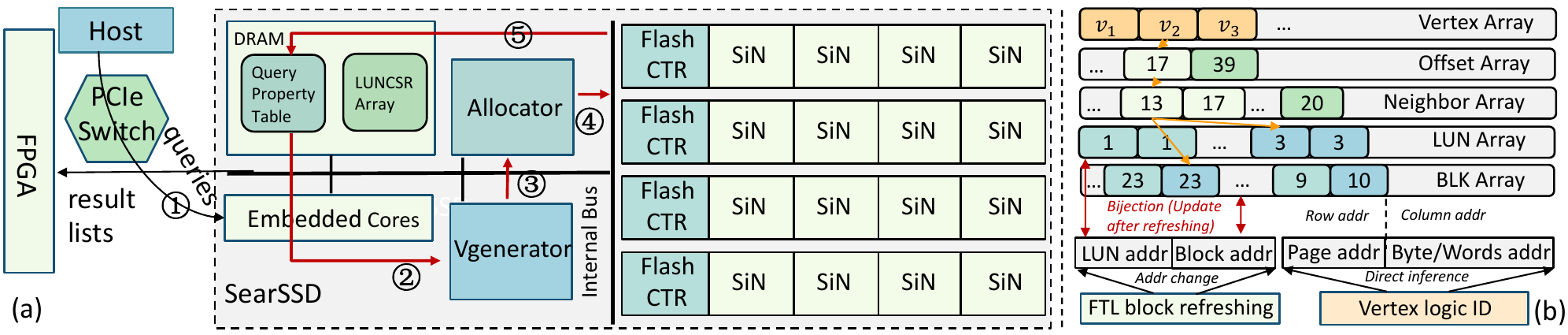}
		\vspace{-0.5em}
		\caption{(a) Overview of \design and overall architecture of \ssd; (b) The new graph format  - LUNCSR with LUN and BLK array.}
		\label{fig:overallnluncsr}
		\vspace{-2em}
	\end{center}	
\end{figure*} 
\begin{figure}[b]
	\begin{center}	
		\vspace{0em}
		\includegraphics[width=0.9\columnwidth]{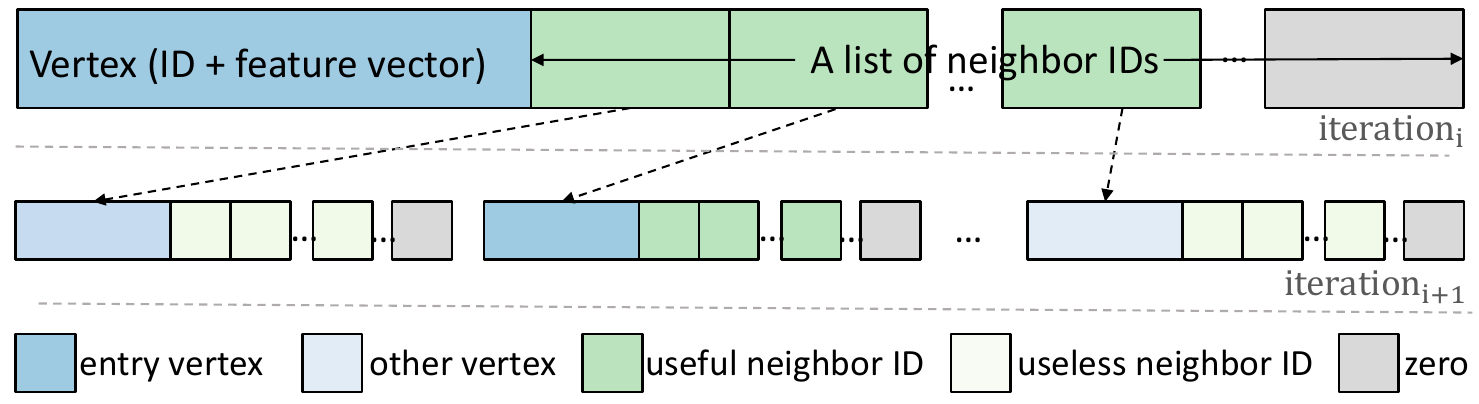}
		\vspace{-0.5em}
		\caption{The inefficient data layout in NDP scenarios.}
		\label{datalayout}
        \vspace{0em}
	\end{center}	
\end{figure} 
\noindent \textbf{Reordering and remapping vertices to improve spatial data locality.} We studied the behavior for the query-wise and batch-wise search of ANNS  at different levels of SSD organization.
Fig.~\ref{accesspattern} (a) illustrates the ratio of the number of page accesses to the length of the searching trace (the number of visited vertices that are computed with the given query), and the ratio of the size of accessed feature vectors to the size of accessed page data of 10 random sampled queries in a batch. The high \textit{\#Accessed pages/\#Length of the trace} ratio and low \textit{Accessed vectors/Accessed page data} ratio indicate that the fine-grained accesses to vertices are scattered among different pages, which means that the page buffer locality is very poor and the access pattern is irregular in the search if the vertices were stored in the order the graph was constructed. This motivates us to reorder the graph vertices and then remap the data to improve spatial data locality in the page buffer, and thus reduce the number of accesses to pages for each query in the search.

\noindent \textbf{Developing LUN-level accelerators and dynamic scheduling to improve temporal data locality.} As shown in Fig.~\ref{accesspattern} (b), during the search of 10 consecutive batches  (with batch size set to 2048) of queries in sift-1b dataset, over 82\% of all the LUNs that store the vertices are accessed in each batch (The vertices are stored in the order of graph construction).
This indicates that the pattern of access to LUNs is highly scattered, and it is possible to explore the LUN-level parallelism and internal bandwidth to speed up the execution. In addition, according to inclusion-exclusion principle~\cite{stanley1986enumerative}, there must be multiple accesses to one LUN when the batch size is larger than the number of total LUNs.
In the original SSD architecture, since the data bus is shared by multiple LUNs, only one LUN of one flash chip in a specific channel can be selected to occupy the bus, thus adopting SSD/channel/chip-level accelerators like ~\cite{mailthody2019deepstore} hinders the operational parallelism. Moreover, reading data from the page buffer to the accelerators outside the NAND Flash chip induces extra $\sim 30 \mu s$ latency.
Hence, based on these observations, we (\romannumeral1) develop LUN-level accelerators based on the existing multi-LUN operations to explore the internal operation parallelism of NAND Flash; and (\romannumeral2) propose a dynamic scheduling mechanism to allocate the queries, whose targeted vertices are in the same LUN,  in one batch to the same LUN based on our LUNCSR graph format, to improve the temporal data locality.

\section{NDSearch Architecture}
\label{sec:micro}

\subsection{Architecture overview}
\label{sec:overall}

This section presents \design, a system that accelerates graph-traversal-based ANNS by leveraging the computational storage device based on SmartSSD \cite{lee2020smartssd} architecture.
Fig.~\ref{fig:overallnluncsr}(a) illustrates the overview of \design, which consists of an FPGA device and a modified SSD device connected by a private PCIe 3.0$\times$4 link. 
As aforementioned, we extract three kernels from the workload of graph-traversal-based ANNS, graph traversal, distance computation, and bitonic sorting. 
We modify the original SSD as \ssd and offload the graph traversal and distance computation kernels to it.
The graph traversal kernel is executed in the embedded cores with additional customized logic (Vgenerator and Allocator) in the SSD.
The distance kernel runs on the Search-in-Nand (SiN) engines where LUN-level accelerators are developed. Each SSD channel consists of four SiN engines. \ssd outputs the result lists of each query to the FPGA. The result list of one query only needs to contain the index of the query, the indices of the query's candidate neighbors, and the scalar distances between the query and the candidates. The feature vectors of query and targeted vertices are ``filtered" by the \ssd to reduce the PCIe bandwidth consumption,
which could be as low as $1/32$ of the data transferred via PCIe link in~\cite{smartSSDKNN}.
The FPGA executes the bitonic sorting kernel and returns the top-k neighbors of each query to the host.

\subsection{Data layout -- LUNCSR}
\label{LUNCSR}

\begin{figure*}[t]
	\begin{center}	
		\includegraphics[width=0.9\textwidth]{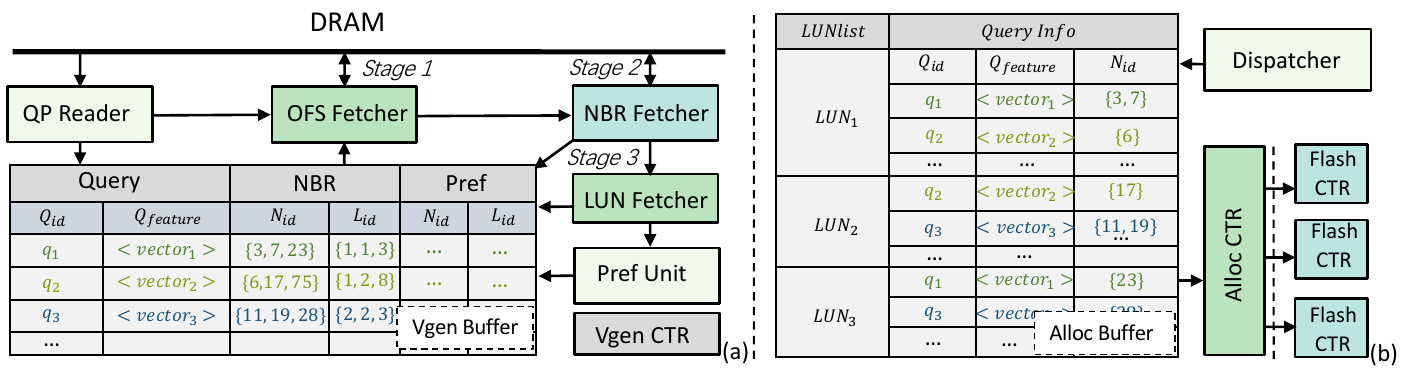}
		\vspace{-0.5em}
		\caption{The detailed architecture of (a) Vgenerator and (b) Allocator.}
		\label{AllocVgen}
	\end{center}	
	\vspace{-1.5em}
\end{figure*} 
DiskANN~\cite{subramanya2019diskann} and HNSW~\cite{malkov2018efficient} adopt the same data  layout such that for each vertex $i$, the feature vector $v_i$ before the IDs of its $\le R$ (i.e., $R = 32$) neighbors and zeros are padded if the degree of a node is smaller than $R$, as shown in Fig.~\ref{datalayout}.  
We argue that this data layout is inefficient for NDP solutions  for two reasons.
First, it wastes space by padding zeros to align the feature vectors and neighbor IDs. 
Second, it fetches unnecessary neighbor IDs that are not used in the search process.
For example, suppose each vertex has a 128-byte feature vector and 32 4-byte neighbor IDs, resulting in a 256-byte slice of data layout.
Then, 16 such slices can fit in one page (assume 4KB page size), which is the minimal access granularity in the NAND array. 
However, during the search, only the neighbor IDs of the closest vertex to the query in each iteration are needed for the next iteration. 
The rest of the neighbor IDs in the page are irrelevant and cause at least 46.9\% storage overhead as shown in Fig.~\ref{datalayout}.
Notice that this data layout is efficient if the graph is stored in CPU main memory or GPU memory, where the access granularity is the cacheline size - 64 bytes, because the smaller granularity decouples the accesses to the feature vectors and neighbor IDs, i.e., accessing the 128-byte feature vector just requires two memory reads without loading neighbor IDs. In addition, the smaller granularity also induces less access to the irrelevant neighbor IDs.
Hence, we found that the compressed sparse row (CSR) is more suitable for the NDP solution since the vertex and neighbor IDs are separately stored. 
We can avoid accessing data that will not be used by the subsequent process under the same page access.

\label{LUNCSR}
CSR is widely used as an efficient format to store graphs.
The original CSR format consists of three one-dimensional arrays: offset, neighbor, and vertex arrays~\cite{csr}. 
The original CSR format does not encode the vertex placement information, which could be critical for NDP.
Thus, we propose LUNCSR, which extends the CSR format by adding two additional arrays, LUN array and block (BLK) array.
The LUN array stores the physical LUN allocation of the vertices, and the BLK array stores each vertex's relative physical block allocation within a LUN.
Both arrays can be indexed by the vertex IDs or the neighbor IDs and updated by FTL when data refreshing occurs. As discussed in Section~\ref{sec:datarefreshing}, NDSearch uses block-level refreshing. 
Fig.\ref{fig:overallnluncsr}(b) shows how FTL updates the LUN and BLK arrays when the block-level refreshing changes the LUN and block addresses. The LUN and BLK arrays are managed in a similar way that FTL manages the mapping table in the traditional SSDs to guarantee the coherency and consistency of the data.
After confirming that no data refreshing happens or the infrequent data refreshing has been done, the specialized Allocator finally generates the physical address of a vertex according to its logical index and the LUNCSR without adopting FTL to do the logical-to-physical address translation.
The page/column address can be directly inferred from the logical index of a vertex since it is not affected by the block-level refreshing. Note that there is no additional memory resources for LUNCSR arrays compared to the standard SSD, where there exists a mapping table for the logical-to-physical address translation. We just transform the mapping table to LUNCSR arrays which can be integrated to \design.
Fig.~\ref{fig:overallnluncsr}(b) also demonstrates how the Allocator indexes the neighbors of $v_2$ in the LUNCSR. 
The arrows illustrate the indexing traces. 
Specifically, the ID (i.e., 2) of $v_2$ points to its first neighbor $v_{13}$ with offset 17. 
The Allocator uses the offset value as the pointer to access the neighbor list of $v_2$ (the length of the neighbor list is the difference between $v_3$'s offset and $v_2$'s offset). 
Then, using the neighbor IDs,  the Allocator can find the neighbors' corresponding LUN and block IDs. The neighbor IDs (also the vertex logic IDs in Fig.~\ref{fig:overallnluncsr}(b)) further indicate the page and column addresses so the physical addresses of each neighbor are finally generated.

\subsection{SearSSD design}
Fig.~\ref{fig:overallnluncsr}(a) depicts the overall architecture of \ssd. The Allocator and Vgenerator are physically implemented on the same die and connected to the internal memory bus of the SSD controller (embedded cores). Only the vertex array in LUNCSR is stored in SiNs. The other arrays are buffered in the internal DRAM or stored in normal NAND flash chips in standard SSD channels which are not illustrated in Fig.~\ref{fig:overallnluncsr} (a).
To support billion-scale ANNS benchmarks, we set the total capacity of SiNs to 512 GB, organized as 32 channels, 4 flash chips per channel, 4 planes per chip, 512 blocks per plane, and 128 pages per block. The page size is 16KB, and we organize two planes as a LUN. Other detailed configurations of \design are shown in Table~\ref{power}.

\subsubsection{Dataflow in \ssd}
The dataflow in \ssd is also shown in Fig.~\ref{fig:overallnluncsr}(a).
\circled{1} A batch of queries is sent from the host to the SSD controller via the PCIe link.
The SSD controller then assigns the initial entry vertex for each query.
A query property table is created in the internal DRAM to store the property of each query (i.e., current searching status, including the ID of the query, the ID of entry vertex in this iteration, the feature vectors of the query, result list and etc.) and maintained by the SSD controller.
\circled{2}  Vgenerator manages to read out the graph information of the entry vertex in LUNCSR format, including offset, LUN IDs, and neighbor IDs. 
The exact neighbor and LUN IDs are generated in Vgenerator and sent to Allocator.
\circled{3} According to the neighbor and LUN IDs from Vgenerator, Allocator allocates the queries and neighbor IDs to different LUNs.
\circled{4} Allocator sends the  queries and physical addresses of the neighbors (candidates) to the corresponding LUN-level accelerators. 
\circled{5} The SiN engines compute the distance between the queries and the vertices stored in the NAND flash chips.
The computed distances are sent back to the SSD controller to update the query property table. 
A new vertex is selected as the updated entry vertex of the next search iteration.
The loop of \circled{2}\circled{3}\circled{4}\circled{5} repeats until the search of the batch of queries terminates.

\subsubsection{Vgenerator}
\label{sec:vgen}
The architecture of the Vgenerator is depicted in Fig.~\ref{AllocVgen}(a). 
The Vgen Buffer is partitioned into three portions: Query, neighbor (NBR), and prefetch (Pref) buffers.
A batch of queries are written into the query buffer.
Then the QP reader reads the IDs of entry vertices in the current search iteration and sends them to OFS Fetcher.
OFS Fetcher, NBR Fetcher, and LUN Fetcher work in a three-stage pipeline to fetch the offset values, the neighbor IDs, and the LUN IDs of the neighbors of the entry vertices, respectively, following the indexing process which is described in Section ~\ref{LUNCSR}. 
Then the NBR Fetcher writes the neighbor IDs into the ``$N_{id}$" fraction of the NBR buffer while LUN Fetcher writes the corresponding LUN IDs into ``$L_{id}$" fraction of the NBR buffer.
The Pref Unit is used to prefetch the neighbors to do the speculative searching, which will be introduced in detail in Section~\ref{sec:scheduling}. 

\subsubsection{Allocator}
\label{sec:alloc}
The architecture of the allocator is illustrated in Fig.~\ref{AllocVgen}(b).
According to the $L_{id}$ in the Vgen Buffer, the Dispatcher gathers the neighbors with the same LUN IDs and the corresponding queries together in the same fraction of Alloc Buffer, which is horizontally partitioned according to LUN IDs.
Then, the Alloc CTR directly generates the physical addresses of all the neighbors using the $N_{id}$s and the corresponding content in LUN/BLK array as described in Section~\ref{LUNCSR}, avoiding the FTL address translation overhead. 
Next, the Alloc CTR sends the data and the physical addresses to the corresponding LUN-level accelerator through Flash CTR for distance computation.

\begin{figure}[t]
    \vspace{0em}
	\begin{center}	
		\includegraphics[width=0.85\columnwidth]{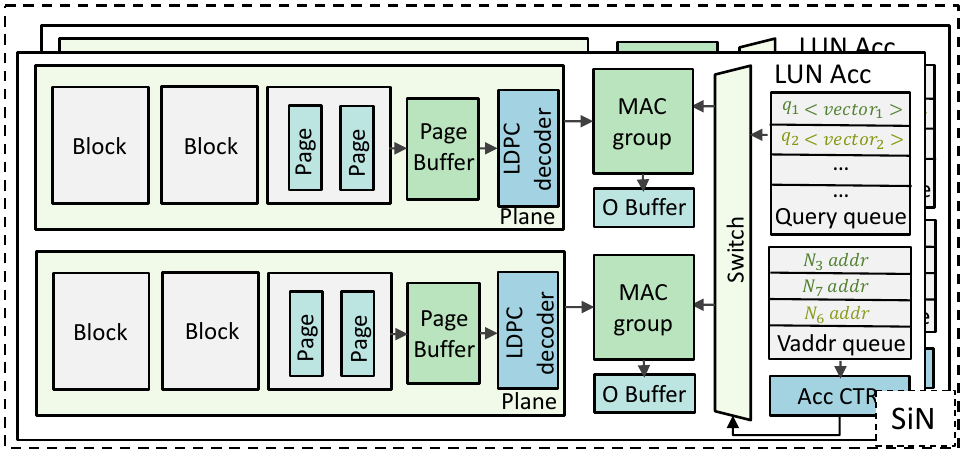}
		\vspace{-0.5em}
		\caption{The architecture of SiN including two LUN accelerators.}
		\label{SiN}
	\end{center}	
	\vspace{-1.5em}
\end{figure} 

\subsubsection{SiN}

Fig.~\ref{SiN} shows the architecture of the SiN engine whose basic unit is a LUN-level accelerator.
One SiN is made up of 2 LUN-level accelerators.
The Flash CTR sends the modified multi-LUN instructions to the SiN to make the LUN-level accelerators in the same SiN process the queries in parallel.
In each LUN-level accelerator, the query queue buffers the feature vectors of queries that are allocated to this LUN, and the Vaddr queue buffers the addresses for the neighbors of each query in the current search iteration.
The Acc CTR sends multi-plane instructions to read vertices from different planes and enables the two multiply-and-accumulate (MAC) groups to work in parallel. The queries are sent to the corresponding MAC group via a switch. The computed distances are temporarily stored in the additional output buffer (O Buffer) for readout. 
Under the premise of 256 LUN-level accelerators in total, we build 2 MACs into each MAC group, whose architecture is based on the adder tree and similar to the design in ~\cite{smartSSDKNN}. The feasibility of the MAC group and its influence on the storage density is discussed in Section~\ref{sec:eva-results}.
\subsubsection{ECC mechanism in SiN}
Error correction code (ECC)~\cite{ber} mechanism is indispensable in SSD, which is used to detect and correct data error induced by the noise and distortion of NAND flash memory cells. We develop the plane-level ECC mechanism in SiN because we should make sure that the accessed feature vectors are corrected before being fed into the MAC group. Hence, as shown in Fig.~\ref{SiN}, a hard-decision decoder~\cite{hard-decision} is put between the page buffer and the MAC group in each plane. We adopt low-density parity-check (LDPC)~\cite{ldpc} codes for ECC, which are initially written into NAND flash memory along with feature vectors. Soft-decision~\cite{soft-decision} LDPC decoding still runs on the FTL layer on the embedded cores and is invoked only if the hard-decision decoding fails. In most cases, implementing hard-decision decoders in SiN is sufficient for the search phase of ANNS. The corresponding evaluation is shown in Section~\ref{sec:eva-results} .


\subsubsection{Multi-LUN operation modification}
\begin{figure}[b]
    \vspace{0em}
     \centering
     \includegraphics[width=0.8\columnwidth]{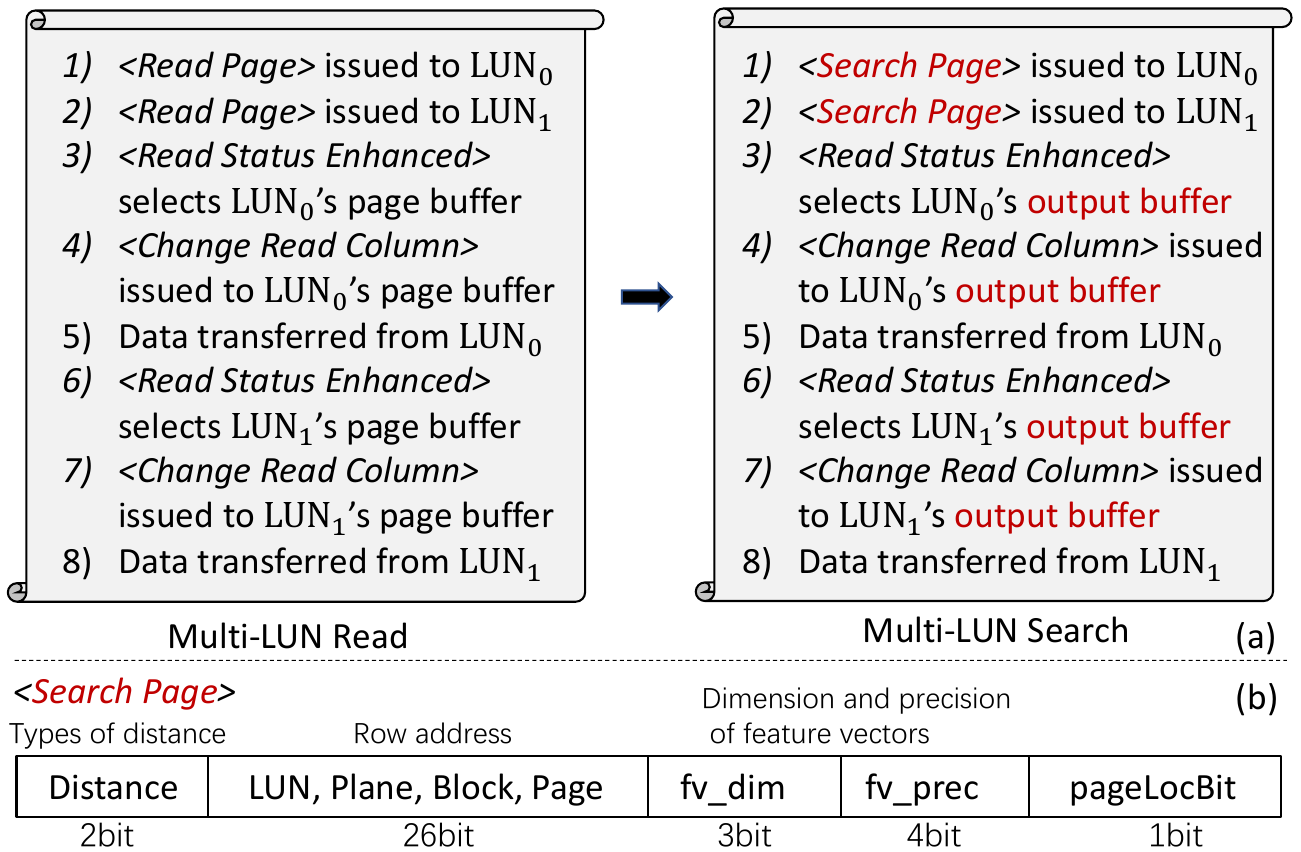}
     				\vspace{-0.5em}
   \caption{(a) The modified workflow of multi-LUN operation; (b) The instruction of search page.}
     \label{multi-LUN}
     \vspace{0em}
\end{figure}
Our multi-LUN search operation is based on the existing multi-LUN read operation in SSD. The typical workflow of multi-LUN read is shown at the left of Fig.~\ref{multi-LUN}(a). We change the \textit{$<Read Page>$} instruction to the specialized \textit{$<Search Page>$} instruction, which is illustrated in Fig.~\ref{multi-LUN}(b). The 2-bit ``Distance" portion indicates which type of distance to compute, like Euclidean distance, angular distance, inner-product distance, and so on. We use 1-bit ``pageLocBit" to illustrate the locality of the page buffer. If ``pageLocBit = 1", there will be two or more queries' candidates located on the selected page. In addition, we change the objects of $<Read Status Enhanced>$ and $<Change Read Column>$ from the page buffer to the output buffer in our design because we only need to transfer the computed distances from each LUN rather than the original feature vectors. The modified workflow of multi-LUN search is shown at the right of Fig.~\ref{multi-LUN}(a).

\section{Processing model}
\label{processingmodel}
Current graph traversal-based ANNS follows the widely adopted processing model in graph analytic~\cite{sundaram2015graphmat}, which consists of two phases - \textit{Scatter} and \textit{Apply}, with three essential operators - \textit{Processing Edge}, \textit{Reduce} and \textit{Apply}.
We found that it is inefficient for \design since it does not consider the dataflow of graph-traversal-based ANNS or the characteristics of the NDP platform. 
Thus, we propose a new processing model of accelerating graph-traversal-based ANNS on \design as shown in Algorithm~\ref{processing model}.
To exploit the parallelism of \design, the model iterate over the LUN list which can be executed in parallel.
To overlap the latency of the dynamic scheduling of graph traversal kernel and the execution of distance computation kernel, we decouple the \textit{Scatter} phase into \textit{Allocating} and \textit{Searching} stages, which will be further discussed in Section~\ref{sec:speculative}.
We also decouple the \textit{Apply} phase into \textit{Gathering} and \textit{Sorting} stages. 
\begin{algorithm}[h]
    \small
    \caption{ANNS Near Data Processing Model}
    \begin{algorithmic}[1]
    \label{processing model}
    \FOR{\textit{i} in range (len(\textit{this batch}))}
    \STATE $q_i.Lid \leftarrow$ Vgenerate (\textit{$q_i.Prop$})
    
    \ENDFOR
    \STATE \textit{LUNlist} $\leftarrow$ Batch-wise dynamic allocating (\textit{this batch})\
    \WHILE{ Searching of \textit{this batch} is not terminated}
    \FOR{\textit{j} in range (len($\#LUNlist$))}
    \FOR{\textit{i} in range (len($LUNlist[j].Q_{id}[i]$))}
    
    \FOR{\textit{k} in range (len($LUNlist[j].q_i.Nid$))} 
    \STATE \textit{ProResult} $\leftarrow$ \textbf{Process Edge} ($LUNlist[j].q_i$, $v_{LUNlist[j].q_i.Nid[k]}$)
    
    \ENDFOR
    \STATE $q_i.tProp$ $\leftarrow$ \textbf{Reduce} ($q_i.Prop$, \textit{ProResult})
    \ENDFOR
    \ENDFOR
    
    \FOR{\textit{i} in range (len(\textit{this batch}))}
    \STATE $q_i.Prop$ $\leftarrow$ \textbf{Apply} ($q_i.tProp$)
    \ENDFOR
    \ENDWHILE
    \RETURN $q_i.top-k$ $\leftarrow$ BitonicSort(for \textit{i} in range (len(\textit{this batch})))
    
    \end{algorithmic}
    \end{algorithm}
    
\noindent \textbf{\textit{Allocating} stage in \textit{Scatter} phase (line 1$\sim$3):} According to the property of each query in the batch, a \textit{LUN look-up table} is generated and the tasks of the search of queries are allocated to the LUN-level accelerators through the batch-wise dynamic allocating method (See Section~\ref{dyn}).

\noindent \textbf{\textit{Searching} stage in \textit{Scatter} phase (line 4$\sim$8):} In the \textit{Searching} stage, separate LUNs work simultaneously with the support of multi-LUN operations. \textit{Process Edge} operator executes the distance computation. \textit{Reduce} operator updates the temporary property of each query, e.g., the result list in the current search iteration. The condition of termination of this stage is determined by the setting in the specific ANNS algorithm.

\noindent \textbf{\textit{Gathering} stage in \textit{Apply} phase (line 9$\sim$10):} After the \textit{Searching} stage, the properties of each query in the batch are updated in the Query Property Table by the \textit{Apply} operator. According to the updated results, the queries that do not meet the termination condition will start the next search iteration which consists of the three stages above.

\noindent \textbf{\textit{Sorting} stage in \textit{Apply} phase (line 11):}
When all queries have met the termination condition, a batch of results lists is sent to the FPGA for sorting. Meanwhile, the allocating stage for the next batch can start.
The top-K nearest neighbors of each query will be selected and sent back to the host.

\section{Two-level Scheduling}
\label{sec:scheduling}
The scheduling involves two levels---static scheduling and dynamic scheduling. 
The static scheduling reorders the vertices offline for a better spatial locality.
Dynamic scheduling processes the graph traverse of a batch of queries at runtime.
\subsection{Static scheduling}
Before reordering the graph, the vertices in the graph are stored according to the constructing order, which is usually random. 
The random order of storing the vertices induces poor data locality because the topology information is not captured.
Prior methods~\cite{coleman2021graph} are either inefficient for breadth-first traversal or incur unacceptable overhead to find a relative good vertices order. Moreover, no prior works consider how to delicately map the reordered data onto the storage devices according to their unique characteristics.
Considering the breadth-first traversal trace in the graph traversal-based ANNS algorithms, we propose our degree ascending breadth-first reordering method based on ~\cite{auroux2015reordering}, \textit{which only requires running once but can achieve near-optimal performance}. We would like to store the neighboring vertices in the graph to the same page in the SSD, to ensure the data locality of page access. However, na\"ively mapping the reordered vertices to the consecutive physical addresses in NAND flash will sacrifice the multi-plane parallelism.
Hence, after reordering, the mapping of the vertices should be coordinated with the restrictions of multi-plane operations of SSD.

\subsubsection{Degree ascending breadth-first traversal reordering}
\begin{figure}[t]
	\begin{center}	
		\includegraphics[width=0.9\columnwidth]{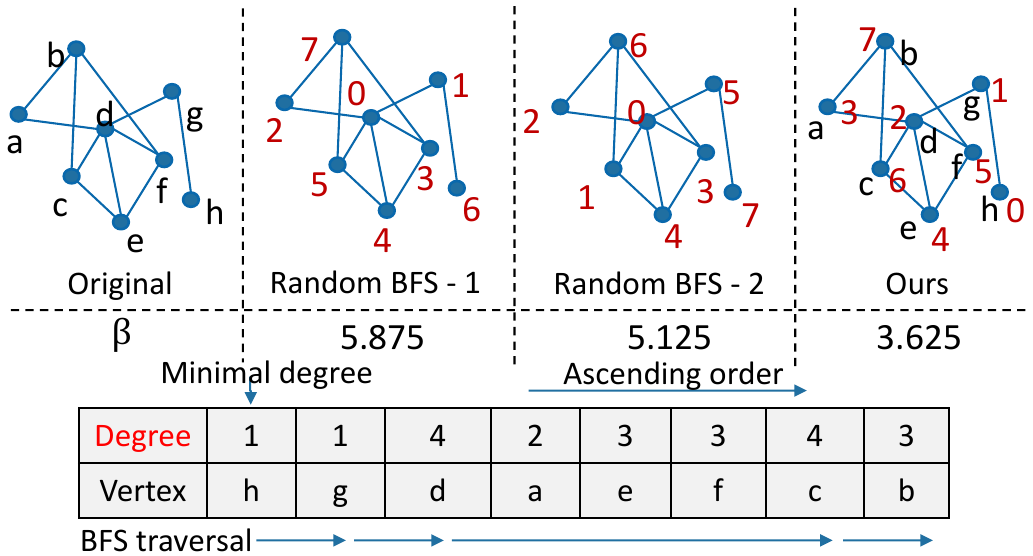}
		\vspace{-0.5em}
		\caption{An example of the comparison between random BFS reordering method and our method.}
		\label{reorder}
	\end{center}	
	\vspace{-2em}
\end{figure}

Let $\mathcal{V} = \{v_1,...,v_n\}$ be the $n$ vertices of a graph $G = (\mathcal{V,E})$ with $|\mathcal{E}|$ edges,  and $f: \mathcal{V}\to \{1,2,...,n\}$ be a reordering function that generate an index for each vertex of the graph.
The goal of the reordering is to find an optimal $f$ to minimize a bandwidth function $\beta (G,f)$ defined as:
\begin{equation}
    \beta (G,f) = \frac{1}{n}\sum_{v \in \mathcal{V} }\mathop{max}\limits_{(i,j)\in \mathcal{E}(v)}|f(i) - f(j)|.
    \label{bandwidth}
\end{equation}
Here, $v$ is each vertex in the graph, and $\mathcal{E}(v)$ generates the indices of all the neighbors of the vertex $v$.
Our objective is to minimize the average vertex bandwidth, $\beta$. 
A small $\beta$ guarantees that the neighbors of each vertex are stored physically close to each other.
Reordering graph vertices to get the minimum $\beta$ has been proved to be an NP-Completeness problem~\cite{NP-C}. Although some prior reordering methods are proposed\cite{caprara2005laying}, the existing randomness in these methods requires running the methods multiple times to approach the optimal reordering result.  
For the example shown in Fig.~\ref{reorder}, if using the prior method in~\cite{auroux2015reordering}, there are 8 choices to select $v_0$ in the original graph. Furthermore, if $v_d$ is selected as $v_0$, there are $5!=120$ choices to renumber $v_a, v_c, v_e, v_f, v_g$ to $v_1 - v_5$ . As the reordering proceeds, the reordering results increase significantly and we should finally select one whose $\beta$ is the smallest.
The overhead of traversing all the possible BFS orders is unacceptable for the huge scale ANNS graphs.
Our reordering method addresses the aforementioned issue of randomness. Specifically, we reorder the vertices based on their degrees in ascending order, which is a deterministic approach rather than a random one. This ordering strategy is motivated by the observation that first randomly labeling the vertices with higher degree (more neighbors) creates difficulty for closely labeling their neighbors (making the indices of the neighbors as close as possible) later, which finally results in a large bandwidth $\beta$.
In contrast, if vertices with lower degrees are reordered first, their neighbors with higher degrees may still remain unnumbered and be easy to be closely labeled, resulting in a smaller $\beta$. In addition, by reordering the vertices with higher degrees later in the process, we can reduce $\beta$ further by placing them closer to their already renumbered neighbors. As a result, our method only needs to be run once to obtain a near-optimal vertex reordering, as illustrated in the final reordered graph in Figure~\ref{reorder}.
Firstly, we select $v_h$ which has the minimal degree - 1, as the root vertex $v_0$. 
Then, the BFS traversal would find $v_g$, and we renumber it to $v_1$. 
After renumbering $v_d$ to $v_2$, we reorder its neighbors according to their degree ascending order. 
In this example, the degrees of $v_a$, $v_c$, $v_e$,$v_f$ and $v_g$, are 3, 4, 3, 3 and 1, respectively. 
Because $v_g$ has been renumbered, we further renumber $v_a$ as $v_3$, $v_e$ as $v_4$, $v_f$ as $v_5$ and $v_c$ as $v_6$. The reordering process will continue until all vertices are renumbered.
Obviously, there is no randomness existing in our method only if the degrees of some neighbors of a vertex are same. Moreover, this example shows that our method can ensure smaller average bandwidth of each vertex by reducing the opportunities that the neighbors are split far away from each other.

\subsubsection{Multi-plane operation}
\begin{figure}[t]
	\begin{center}	
		\includegraphics[width=0.9\columnwidth]{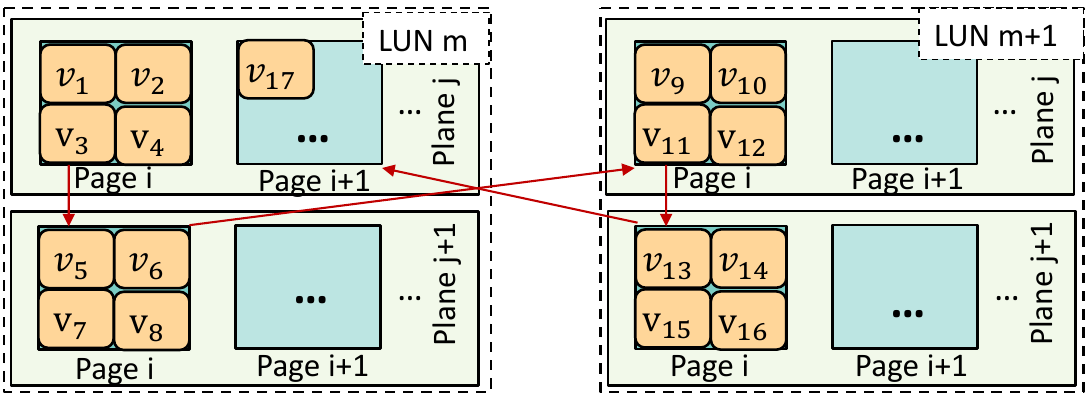}
		\vspace{-0.5em}
		\caption{Mapping of vertices restricted by multi-plane addressing.}
		\label{mapping}
	\end{center}	
	\vspace{-1.5em}
\end{figure}

To exploit the parallelism of  multi-plane operations, we should map the reordered vertices under the restrictions of multi-plane addressing.  
There are two restrictions applied to the multi-plane address when executing a multi-plane command sequence on a particular LUN: (\romannumeral1) The plane address bits shall be distinct from any other multi-plane operation in the multi-plane command sequence; (\romannumeral2) The page/LUN address shall be the same as any other multi-plane operations in the multi-plane command sequence. 
Hence, our mapping strategy is that we first map the reordered vertices in one page of a plane to one LUN, e.g., $page_i$ in $plane_j$, to maximize the data locality in one page. 
Then, we choose the same $page_i$ in another $plane_{j+1}$ in the same LUN. 
Next, we iteratively perform the mapping with the aforementioned process on different LUNs. 
If we have selected all LUNs, we go back to the first LUN and select a different page number for the subsequent vertices.
As illustrated in Fig.~\ref{mapping}, the arrow shows the path of mapping. 

\label{sec:staticoverhead}
The offline static scheduling could take several hours, depending on the specific constructed graph. Our reordering process operates independently of the SSD's organization, thus eliminating the need for it to be repeated when switching to a different SSD. However, the mapping process must be re-conducted as it relies on the internal architecture of the SSD. In addition, we also consider the the influence of data refreshing on our static scheduling. As aforementioned, we adopt block-level data refreshing in \ssd. Obviously, our degree ascending breadth-first traversal reordering is not affected because we reorder the vertices to improve the page-level data locality. To avoid degrading the parallelism of multi-plane operation, which is achieved by our mapping strategy, we make the block-level data refreshing happen within planes instead of arbitrary locations in the SSD.

\subsection{Dynamic scheduling}
\label{dyn}
Our dynamic scheduling, which is executed by Vgenerator and Allocator, aims to efficiently allocate queries to the corresponding LUN-level accelerators to improve temporal data locality in each LUN and overlap the latency between search iterations within one batch of queries. It consists of batch-wise dynamic allocating and speculative searching.
\subsubsection{Batch-wise dynamic allocating} 

    
    
This technique allocates queries with the same targeted LUNs to the corresponding LUN-level accelerators at once. The implementation of the batch-wise dynamic allocating is suggested in the architecture design of Vgenerator and Allocator in Section~\ref{sec:vgen} and Section~\ref{sec:alloc}
An example is illustrated in Fig.~\ref{AllocVgen}, where $q_1$ is sent to $LUN_1$ and $LUN_3$, $q_2$ is sent to $LUN_1$ and $LUN_2$, and $q_3$ is sent to $LUN_2$ and $LUN_3$ to be computed distances with the corresponding neighbors.
Note that one query can be allocated to different LUN-level accelerators. The allocated queries in each LUN are further assigned to the corresponding planes in a similar way to allocating queries to LUNs. In this way, the pages consisting of the candidate neighbors of different queries only need to be loaded once from the plane and the temporal data locality in each LUN is fully exploited.


\subsubsection{Speculative searching}
\label{sec:speculative}
\begin{figure}[t]
	\begin{center}	
        \vspace{0em}
		\includegraphics[width=0.9\columnwidth]{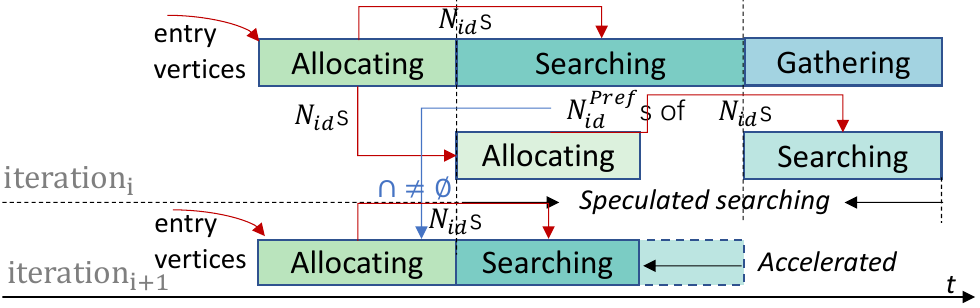}
		\vspace{-0.5em}
		\caption{The mechanism of speculative searching.}
		\label{speculative}
	\end{center}	
	\vspace{-1.5em}
\end{figure}

As aforementioned in Section~\ref{processingmodel}, a common search iteration consists of three sequential stages: \textit{Allocating} stage, \textit{Searching} stage, and \textit{Gathering} stage as illustrated in Fig.~\ref{speculative}. The \textit{Allocating} stage of the next search iteration usually requires the updated results of \textit{Gathering} stage of iteration$_i$ to determine the entry vertices in iteration$_{i+1}$. The decoupling of the three stages provides the opportunity to overlap the latency of the \textit{Allocating} stage  and \textit{Searching} stage. As shown in Fig.~\ref{speculative}, we develop a speculative searching mechanism underlying the processing model  based on \textit{the observation that the second-order neighbors of the entry vertex in the current iteration are the potential candidates to access in the next search iteration.} Thus, the second-order neighbors are highly likely to be accessed in the next iteration. According to this observation, in the search iteration$_i$, when the \textit{Allocating} stage is done, we can get the neighbor IDs - $N_{id}$s - of each entry vertex in the current iteration. When the \textit{Searching} stage of this iteration begins, we start the speculative searching for the next iteration - iteration$_{i+1}$ by launching the speculative \textit{Allocating} stage in iteration$_i$. The Pref Unit fetches the neighbors of each entry vertex in iteration$_i$ and generates the corresponding IDs of some second-order neighbors of each vertex-  $N^{Pref}_{id}$s for each entry vertex in iteration$_i$. Considering the number of second-order neighbors is usually larger than that of the first-order neighbors of each entry vertex, the Pref Unit selects the second-order neighbors that have more connections with the first-order neighbors. The $N^{Pref}_{id}$s are stored in the Pref buffer of Vgen Buffer.  When the \textit{Searching} stage of iteration$_i$ ends and the \textit{Gathering} stage of iteration$_i$ starts, the speculative \textit{Searching} stage launches to compute the distances between the queries with their prefetched neighbors. Then, for a query, if there is an overlap between its $N_{id}$s in iteration$_{i+1}$ and $N^{Pref}_{id}$s in iteration$_i$ ($N^{Pref}_{id} \cap N_{id} \neq \varnothing $), the corresponding speculative searching results can be used. In this way, the \textit{Searching} stage of iteration$_{i+1}$ can be accelerated. Note that if the speculative \textit{Allocating} stage does not complete when the non-speculative \textit{Searching} stage ends in iteration$_i$, the speculative \textit{Allocating} will be forcibly terminated. Hence, the latency of the speculative searching can be entirely overlapped. 
\begin{figure*}[t]
	\begin{center}	
		\includegraphics[width=\textwidth]{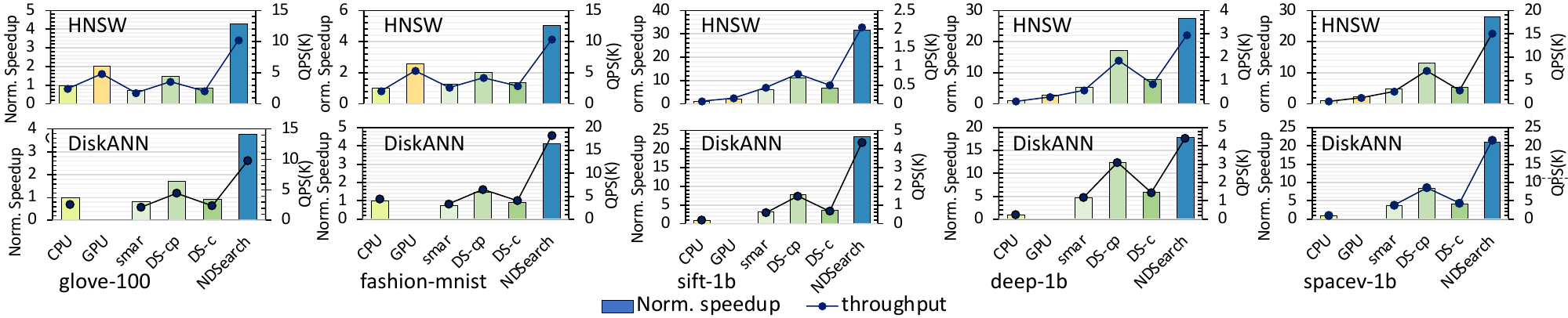}
		\vspace{-1.5em}
		\caption{Speedup normalized to CPU (shown in the histogram) and throughput (shown in the line chart) comparison on various platforms. We measure the throughput by processing a batch (2048) of queries with the same memory trace on each benchmark.}
		\label{speedup}
	\end{center}	
	\vspace{-1em}
\end{figure*} 

\section{Evaluation}
\label{sec:eval}
\subsection{Experiment methodology}

\noindent \textbf{Benchmarks and datasets.} We evaluate  \design on two typical graph-traversal-based ANNS algorithms: HNSW\yt{~\cite{malkov2018efficient}} and DiskANN~\cite{subramanya2019diskann}. 
For HNSW, we select   \textit{hnswlib}~\cite{hnswlib} and \textit{cuhnsw}~\cite{cuhnsw} to compare \design with CPU and GPU platforms, respectively. DiskANN has only one implementation~\cite{diskann} that is designed for the CPU platform, regarding the main memory as the ``cache" of the SSD in its algorithm.
Each algorithm is customized and evaluated with 5 datasets: glove-100~\cite{pennington2014glove}, fashion-mnist~\cite{xiao2017fashion}, sift-1b~\cite{sift}, deep-1b~\cite{babenko2016efficient} and spacev-1b~\cite{spacev}. 
We tune the parameters of HNSW and DiskANN with the recall@10 reaching 95\%, 95\%, 94\%, 93\%, 90\% on glove-100, fashion-mnist, sift-1b, deep-1b, and spacev-1b, respectively, to construct the reasonable graphs.

\noindent \textbf{Experimental setup.}
For the SSD part - \ssd, we build an in-house simulator of \ssd based on SSD-Sim~\cite{ssdsim2,ssdsim}, which is a memory trace-based and cycle-level simulator.
We model the behavior and set parameters of SSD based on Samsung 983 DCT 1.92T~\cite{pm983}. The SSD internal DRAM is set to 4GB.
We model the additional buffers and queues in \ssd using CACTI 6.5~\cite{cacti} with 32nm technology. 
We implement and synthesize the digital logic circuits at the 32nm technology node with 800MHz using Synopsys Design Compiler. 
For the FPGA part, which is not the focus of this paper, we directly adopt the similar implementation of bitonic sort kernel in ~\cite{salamat2021nascent}. 
The \ssd and FPGA are connected through a PCIe 3.0 $\times$ 4 bus. We use 2 Intel Xeon Gold 6254 CPUs running at 3.1 GHz with 24GB DRAM (same as GPU memory) as the CPU baseline and a NVIDIA Titan RTX with 24GB VRAM as the GPU baseline. As aforementioned, for HNSW, when the size of the dataset exceeds the memory capacity, we split the dataset into several smaller shards through k-means and load a few shards to the memory from storage to run the algorithm on CPU or GPU.
To compare \design with previous NDP architectures, We build a DeepStore accelerator as a baseline with different levels of accelerators~\cite{mailthody2019deepstore} using the same budget with \design and a SmartSSD-only design like ~\cite{smartSSDKNN} as another baseline.

\noindent 
\textbf{Simulation method.}
We firstly run the construction of ANNS graph on CPU and GPU to get the adjacency information of the graph. Then, we reorder the graph using our reordering method and get the LUNCSR. To generate the memory trace during the search phase, we hack the code of HNSW and DiskANN, initialize the entry vertex for each query and run the search phase of the algorithm on CPU or GPU to get the memory traces, which illustrate the index sequences of the accessed vertices for each query. After that, we feed the traces as input to our trace-driven simulator. 

\begin{figure}[t]
	\begin{center}	
		\vspace{0em}
		\includegraphics[width=\columnwidth]{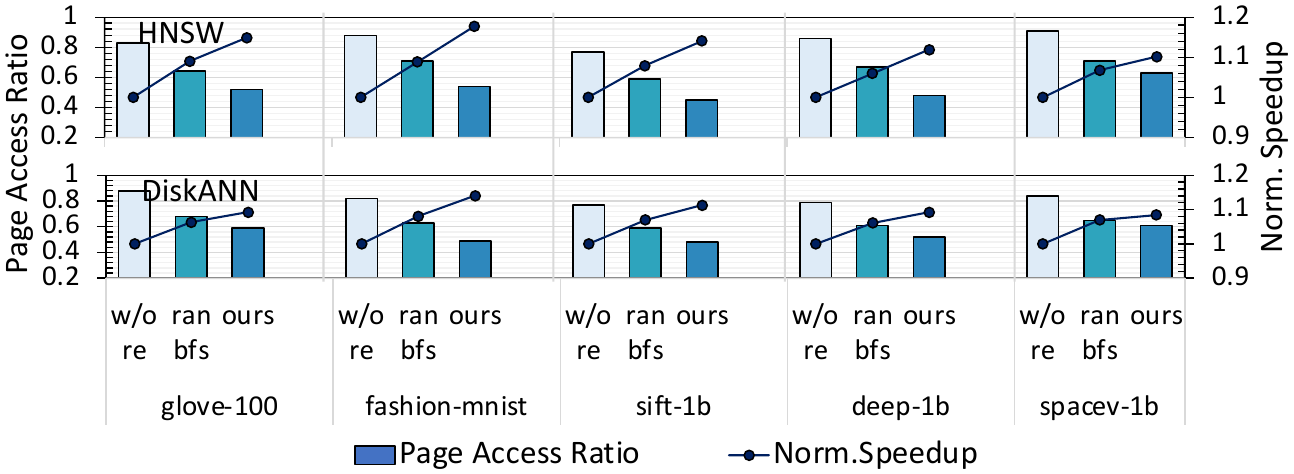}
		\vspace{-1.5em}
		\caption{Evaluation of static scheduling of \design. Speedup is normalized to \design without any reordering.}
		\vspace{-1.5em}
		\label{mapresult}
	\end{center}	
\end{figure}



\subsection{Results}
\label{sec:eva-results}
\noindent \textbf{Performance.} Fig.~\ref{speedup} shows the results of throughput (Query Per Second) and the normalized speedup that \design achieves over CPU, GPU, and DeepStore. For DeepStore, we build a channel-level accelerator (DS-c) and chip-level accelerator (DS-cp) as the baselines. The default batch size is set to 2048. Thanks to the high internal parallelism, GPU, DeepStore, and \design can achieve much better performance compared to CPU when processing a large batch of queries. 
In terms of large datasets, the constructed graph of sift-1b, deep-1b, and spacev-1b could consume more than 500 GBs of memory which exceeds the capacity of VRAM in GPU.
Hence, GPU should load a few shards of the data from the SSD several times through the PCIe link. 
We can observe that DS-c, DS-cp, and \design all outperform the SmartSSD-only design because the SmartSSD-only design does not explore the internal bandwidth and parallelism of the SSD device. Especially, \design achieves up to 7.44$\times$ speedup over the SmartSSD-only design when running DiskANN on sift-1b.
It is interesting that DS-cp achieves higher speedup than DS-c on these benchmarks, which differs from the conclusion in ~\cite{mailthody2019deepstore}. 
This is because the workload of graph traversal-based ANNS does not include compute-intensive operations as shown in Fig.~\ref{Roofline} like the neural network evaluated in the DeepStore paper. 
Hence, the limited computing resources will not be the bottleneck of chip-level accelerators. 
The chip-level accelerators enable DS-cp to process the graph traverse and distance computation locally near the chip and thus DS-cp performs better than DS-c. 
Compared with DS-cp, \design develops LUN-level accelerators with even higher parallelism.
Besides, we utilize and modify the flash chip's internal multi-LUN and multi-plane operations to support access to vertex vectors and distance computation. 
The result shows that \design can achieve up to 2.81$\times$ and 2.94$\times$  speedup over DS-cp when running HNSW and DiskANN, respectively. Hence, we can conclude that more fine-grained accelerators at the LUN level are more efficient for ANNS.

It is not difficult to find that \design demonstrates a higher speedup over CPU/GPU when running HNSW/DiskANN on sift-1b/deep-1b are better than those on glove-100/fashion-mnist. 
This is because the glove-100 and fashion-mnist are much smaller than other datasets.
The constructed graphs of these two datasets are able to fit into the main memory of CPU and the memory of the GPU so that CPU/GPU only needs to load the data from the SSD once. Common NDP designs like SmartSSD, DS-c, and DS-cp can hardly outperform CPU/GPU when running ANNS on small datasets because their advantages over CPU/GPU mainly come from reducing SSD I/O read. However, \design further exploits the SSD internal bandwidth with LUN-level accelerators and develops efficient scheduling schemes to increase the searching parallelism. Hence, \design can still achieve up to 5.06$\times$ and 2.12$\times$ speedup over CPU and GPU, respectively.



\begin{figure}[t]
    \vspace{0em}
	\begin{center}	
		\includegraphics[width=\columnwidth]{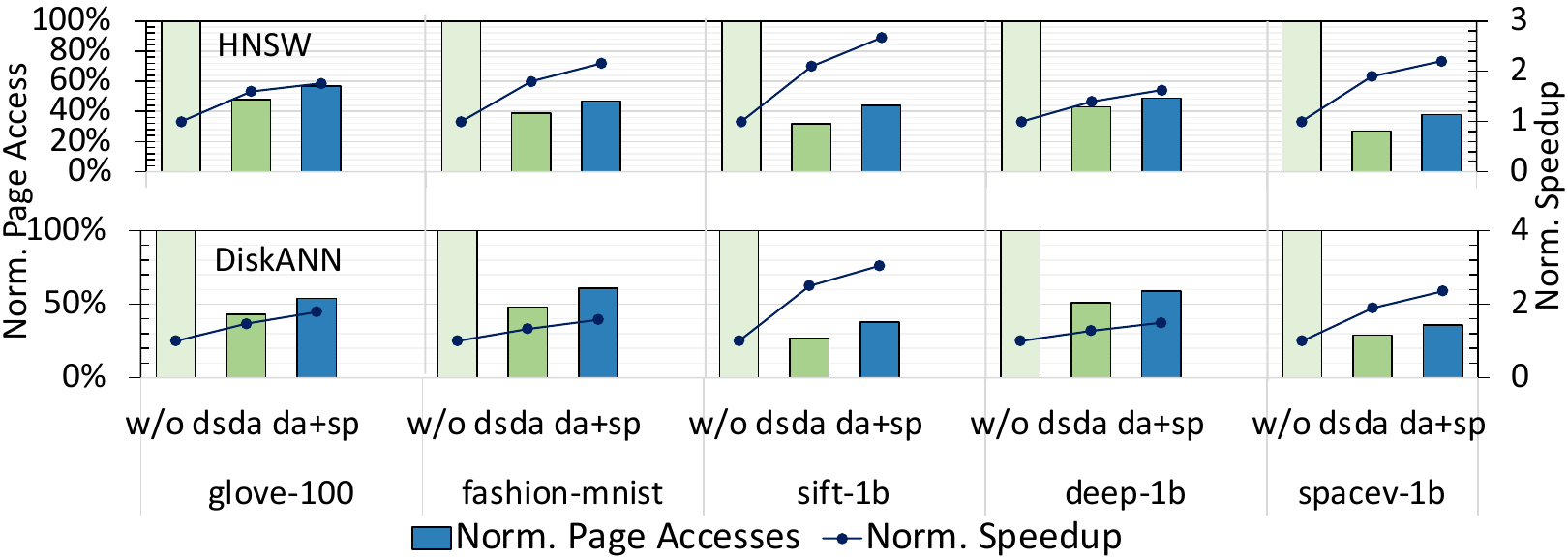}
		\vspace{-1.5em}
		\caption{Evaluation of dynamic scheduling. Page access and speedup are normalized to \design without dynamic scheduling. }
		\label{dsresult}
	\end{center}	
	\vspace{-1.5em}
\end{figure} 
\noindent \textbf{Scheduling.}
To evaluate the performance of our reordering method applied to \ssd, we define a new metric, page access ratio, which equals the ratio of the number of page accesses to the length of the searching trace of a query. 
A higher page access ratio reflects that each page access returns fewer requested/prefetched vertices, thus indicating poor spatial data locality.
We set the page size to 16 KB in this experiment and compute the average page access ratio of a batch (2048) of queries on each benchmark. After applying our reordering method, the bandwidth $\beta$ (defined in Section~\ref{sec:scheduling}) decreases, which means that the neighbors of each vertex are stored closer to each other. Thus, the neighbors of each vertex are more likely to be stored in the same page, which implicitly reduces the number of page accesses. 
Fig.~\ref{mapresult} shows the comparison of different settings: without reordering (w/o re), with random BFS reordering (ran bfs), and ours. The three settings are configured on \design with dynamic scheduling, respectively.
The results indicate that after reordering the vertices and mapping them under the restrictions of multi-plane operation, our method reduces the page access ratio by up to  38\% and induces up to 1.17$\times$ speedup over the baseline without reordering.

We also evaluate the contribution of our dynamic scheduling by configuring \design with three different settings: without dynamic scheduling (w/o ds), with dynamic allocating(da) and with dynamic allocating and speculative searching (da+sp).  The three settings are configured on \design with static scheduling, respectively. In the setting of \design without dynamic scheduling, each query is just allocated to LUNs sequentially according to the addresses of its targeted vertices that may be flushed and need to be read from the NAND arrays again by another query later. By contrast, with dynamic allocating, different queries that are allocated to the same LUN are potentially able to share the same page access as much as possible. Hence, redundant re-allocating and extra page accesses are avoided. As illustrated in Fig.~\ref{dsresult}, dynamic allocating can help reduce the page accesses by up to 73\% and induce up to 2.67$\times$ speedup. With speculative searching, the page accesses increase because over half of speculated results are not selected, which leads to extra accesses to the second-order neighbors of the queries. However, speculative searching can still further induce up to 1.27$\times$ speedup.


\noindent \textbf{Ablation study.}
\begin{figure}[t]
	\begin{center}	
		\vspace{0em}
		\includegraphics[width=\columnwidth]{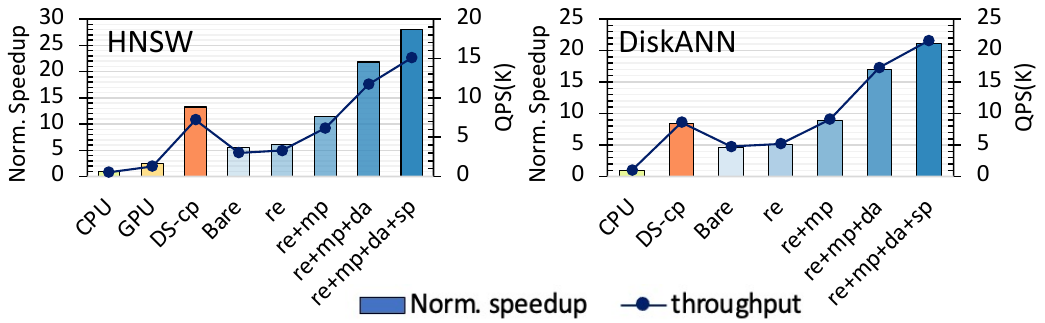}
		\vspace{-1em}
		\caption{The ablation study of our proposed techniques on \design. The experiments are run on spacev-1b dataset.}
		\vspace{-1em}
		\label{ablationstudy}
	\end{center}	
\end{figure} 
\label{sec:ablationstudy}
We conduct the ablation study of the proposed techniques, including degree ascending BFS reordering (re), multi-plane operation mapping (mp), dynamic allocating (da) and speculative searching (sp) on spacev-1b as shown in Fig.~\ref{ablationstudy}. Even without any optimizations, the bare  machine of \design (Bare) can still achieve over 4$\times$ speedup over CPU because 1) data transfer via the PCIe is eliminated; and 2) \design does not need frequent access DRAM to fetch the vertices as CPU does to its main memory. Without the dynamic allocating, \design can hardly beat DS-cp though \design achieves larger parallelism and internal bandwidth due to multi-LUN/plane operations. This is because there are redundant operations on the LUN-level accelerators and we actually implement dynamic allocating on DS-cp to maximize its hardware utilization. With all the proposed scheduling techniques, \design can fully exploit the potential of our hardware design and further achieve 4.1$\times$ performance improvement compared to bare \design, showing the benefits of our software-hardware co-design solution.

\begin{figure}[t]
	\begin{center}	
		\vspace{0em}\includegraphics[width=0.9\columnwidth]
        {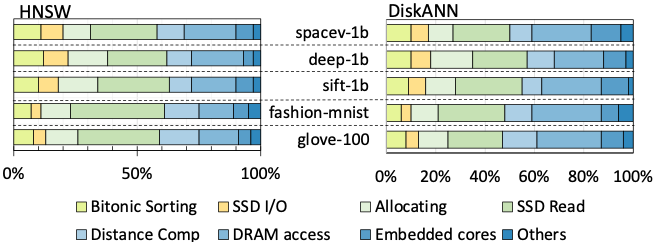}
		\vspace{-0.5em}
		\caption{Breakdown of execution time of \design.}
		\label{breakdown}
	\end{center}	
	\vspace{-2em}
\end{figure} 
\noindent \textbf{Overhead analysis.} 
As illustrated  in Fig.~\ref{breakdown}, NAND read occupies the largest proportion ($24\% - 38\%$) of the entire execution time of \design due to the frequent access to feature vectors stored in NAND flash chips. However, compared to the CPU+SSD system, the proportion of SSD I/O read is reduced from $\sim$70\% (as shown in Fig.~\ref{AccessSSD}) to $\sim$6\% thanks to the ``filtering" of \ssd. The latency of the bitonic kernel on FPGA only accounts for at most 12\% of the overall latency. The ``Allocating" part captures the overhead of dynamic scheduling, specifically the batch-wise dynamic allocating. The speculative searching overhead is overlapped by the distance computation and SSD read in the non-speculative searching stage as discussed in Section~\ref{sec:speculative}. We can also observe that running DiskANN requires more DRAM access and execution of  embedded cores but fewer SSD reads than HNSW thanks to using the internal DRAM of SSD to cache some hot feature vectors. Generally, DRAM access and execution of embedded cores take 20$\%- 35\%$ of the total execution time to maintain the FTL, access LUNCSR, and buffer the temporary results. \label{sec:overhead}

\noindent \textbf{ECC and endurance.}
\begin{figure}[t]
	\begin{center}	
		\includegraphics[width=0.9\columnwidth]{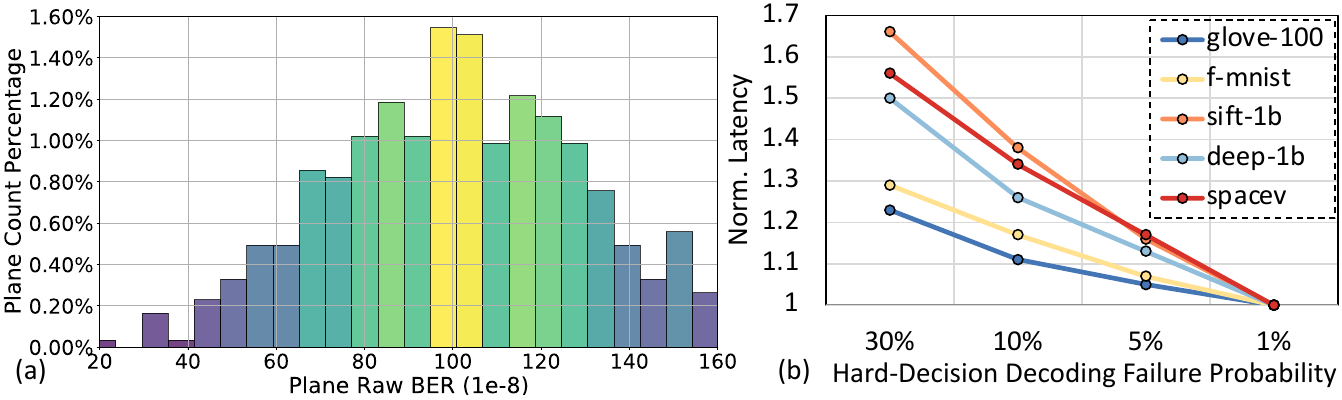}
		\vspace{-0.5em}
		\caption{(a) Plane-level distribution of raw bit error rate; (b) Normalized latency of running  HNSW workloads with different hard-decision decoding failure probabilities.}
		\label{ECC}
	\end{center}	
	\vspace{-1em}
\end{figure} 
We generate the raw bit error rate (BER) statistics of 512 planes in \ssd following the measured memory BER distribution in ~\cite{ber} as shown in Fig.~\ref{ECC}(a). We set the average BER to $10^{-6}$, which is the typical BER value of current advanced NAND flash. We also consider the fact that the probability of hard-decision LDPC decoding increases since flash memory cell storage reliability gradually degrades. As reported in ~\cite{ber}, even at the mid-late lifetime of flash memory, a hard-decision LDPC decoder can still have a low failure probability ($1\%$, which is also our default case). We evaluate our ECC mechanism in worse scenarios with the hard-decision decoding failure probability set to $30\%$, $10\%$, $5\%$ and $1\%$, respectively. When the hard-decision decoding fails, the soft-decision starts to work on FTL. Following the logistics of fault injection~\cite{faultinjection}, we ``inject" the raw BER and the hard-decision decoding failure probability into our simulation environment. As illustrated in Fig.~\ref{ECC}(b), In the worst cases where the hard-decision decoding failure probability is 30\%, \design is slowed down by between 1.23$\times$ and 1.66$\times$. The slowdown mainly comes from the extra latency of executing soft-decision LDPC ($\sim$ 10 $\mu s$) on FTL and pausing the search iteration. We generally conclude that the plane-level hard-decision LDPC decoder is sufficient in most cases.

\begin{figure}[t]
    \vspace{0em}
	\begin{center}	
	   \vspace{0em}
		\includegraphics[width=0.9\columnwidth]{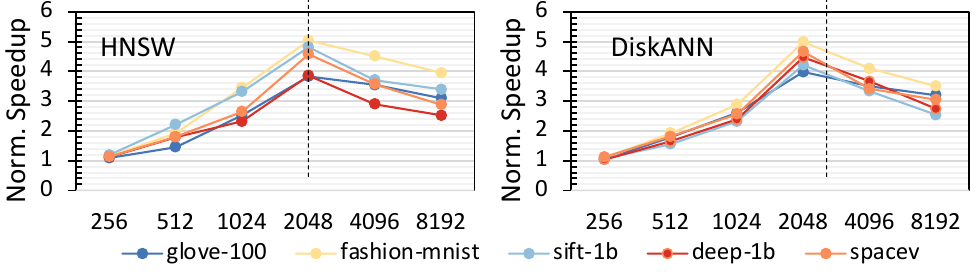}
		\vspace{-0.5em}
		\caption{Speedup normalized to DS-cp with different batch sizes.}
		\label{batchsize}
        \vspace{-2em}
	\end{center}	
	
\end{figure} 

\noindent \textbf{Batch size.}
We explore the impact of batch size on performance. 
As illustrated in Fig.~\ref{batchsize}, we set the batch size ranging from 256 to 8192 to evaluate \design against DS-cp. 
We observe that when the batch size is set to 256, \design only has a marginal advantage over DS-cp. The parallelism of  relatively fine-grained LUN accelerators cannot be fully exploited when the batch size is small.
The irregular access pattern of queries could allocate queries to a few LUN accelerators.
In this situation, the advantage of \design is negligible compared to directly processing queries in relatively coarse-grained chip-level accelerators.
However, as the batch size increases, \design gains a significant advantage because each LUN has a heavier workload so that the LUN-level parallelism can be fully exploited, and the queries are allocated to most LUNs. 
The overhead of gathering data from the flash chip to the chip-level accelerators (as aforementioned, only one LUN can be selected when reading data from a flash chip) limited the performance of DS-cp. When the batch size increases to 4196, the speedup of \design begins to decrease because  the batch have to be split into two or more sub-batches to process, due to the limited resources setting of our design considering the power budget.


\begin{figure}[t]
	\begin{center}	
		\vspace{-0.5em}\includegraphics[width=0.85\columnwidth]{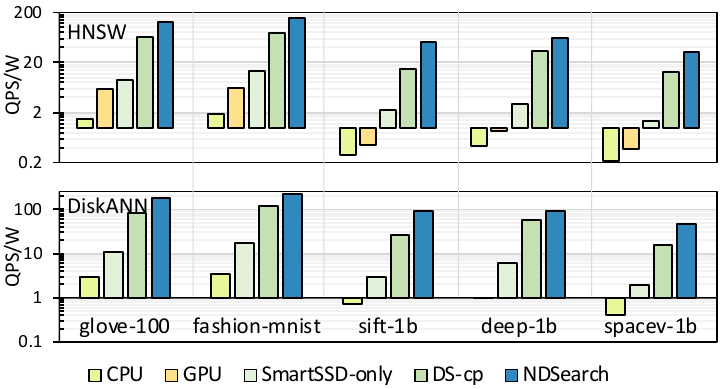}
		\vspace{-0.5em}
		\caption{Comparison of energy efficiency over various designs.}
		\label{energy-efficiency}
	\end{center}	
	\vspace{-2em}
\end{figure} 

\begin{table}[t]
\footnotesize
\centering	
\vspace{0em}
\caption{Power and Area Breakdown of \ssd}
\vspace{-1em}
\begin{tabular}{|c|c|c|c|c|}
\hline
\hline
\  & \textbf{Config.} & \textbf{Num.}  & \textbf{Power} & \textbf{Area}\\
\hline
MAC group & 2 MACs & 512 & 1.95 W & 15.04 $mm^2$ \\
\hline
Vgen Buffer & 2MB &  1 & 1.71 W & 3.18 $mm^2$ \\
\hline
Alloc Buffer & 6MB & 1 & 4.57 W & 8.53 $mm^2$\\
\hline
Query Queue & 24KB &  256 & 5.84 W & 9.76 $mm^2$\\
\hline
Vaddr Queue & 3KB &  256 & 0.87 W & 1.47 $mm^2$\\
\hline
Output Buffer & 1KB & 512 & 0.56W & 1.12 $mm^2$\\
\hline
ECC Decoder & LDPC &  1024 & 1.18W & 2.84 $mm^2$\\
\hline
Ctr circuits & - & - & 2.14W & 1.15 $mm^2$\\
\hline
\hline
Overall & - &  - & 18.82W & 43.09 $mm^2$ \\
\hline
\end{tabular}
\vspace{-2em}
\label{power}
\end{table}

\noindent \textbf{Power budget and Energy Efficiency.}
\label{powerbudget}
The power budget of \ssd is limited by the PCIe interface, which provides $\sim$55W budget ~\cite{mailthody2019deepstore} for \ssd's design. Table~\ref{power} shows the power breakdown in \ssd's design. Considering that the bitonic sorting kernel consumes 7.5W on the FPGA, the total power of \design is 26.32W, which is within the power budget. Fig. ~\ref{energy-efficiency} shows the comparison of energy efficiency with various platforms. Basically, the less the data transfer of feature vectors is and the higher internal parallelism the design achieves, the higher energy efficiency is reached. The SmartSSD-only design avoids the data transfer passing through the host CPU. Compared to the SmartSSD-only design, a large amount of data transfer via SSD I/O is avoided in \design, and  our \design further reduces the data movement from NAND flash chips by the in-LUN computing. Overall, \design achieves up to 178.68$\times$, 120.87$\times$, 30.06$\times$ and 3.48$\times$ higher energy efficiency than CPU, GPU, the SmartSSD-only design, and DS-cp, respectively.

\noindent \textbf{Area and storage density.} 
\label{sec:area}
Area breakdown of \ssd is also illustrated in Table~\ref{power}. With the technology node of 32nm, the total area of the customized logic in our design is 43.09 $mm^2$ while the customized logic in DS-cp and DS-c takes 236.8 $mm^2$ and 320 $mm^2$~\cite{mailthody2019deepstore}, respectively. The area of \design is 82\% and 87\% less than that of DS-cp and DS-c, respectively. SmartSSD takes around 800 $mm^2$ to implement all the logic except for the SSD area~\cite{smartSSDKNN}. 
We estimate the storage density of Samsung 983 DCT~\cite{pm983}, which adopts V-NAND MLC, as $6Gb/mm^2$. After adding the specialized logic inside the SSD, the storage density is reduced to (capacity of \ssd, 512GB) $\times$ 8b/B / ((capacity of \ssd,  512GB) $\times$ 8b/B / 6Gb/$mm^2$ + 43.9 $mm^2$))  = 5.64 Gb/$mm^2$, which is acceptable with only 6\% density degradation.

\section{Discussion}
\subsection{Evaluation on other graph-traversal-based ANNS}
\label{new_alg_eval}
\begin{figure}[t]
	\begin{center}	
		\includegraphics[width=0.8\columnwidth]{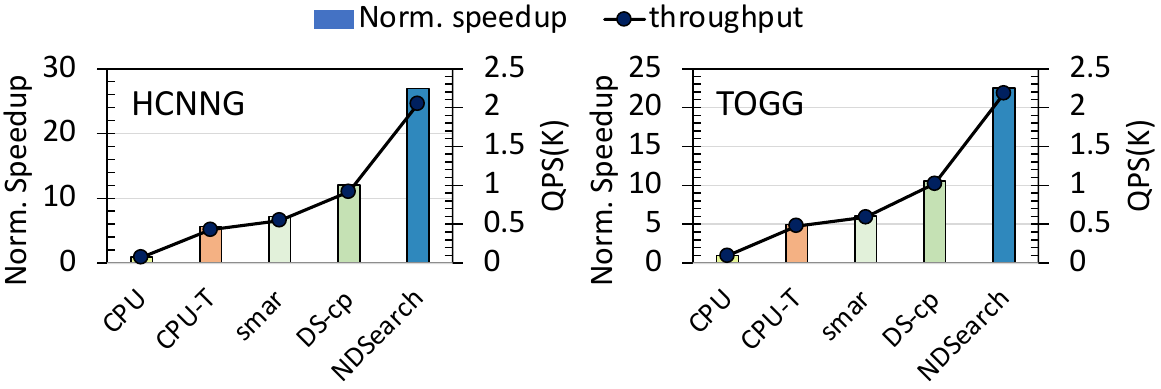}
		\vspace{-0.5em}
		\caption{The normalized speedup (to CPU) and throughput of HCNNG~\cite{hcnng} and TOGG~\cite{togg} on sift-1b on various platforms.}
		\label{HCNNG}
	\end{center}	
	\vspace{-1.5em}
\end{figure}

Besides HNSW and DiskANN, there are some other emerging graph-traversal-based ANNS algorithms like HCNNG~\cite{hcnng} and TOGG~\cite{togg}. Based on the breadth-first search kernel, these algorithms further optimize the search phase by guiding the direction of the query in the vector space.  We construct the graph with HCNNG and TOGG on sift-1b, tuning the recall rate@10 to 93\% and evaluate the search phases on various platforms as shown in Fig.~\ref{HCNNG}. We change the control logic of the embedded cores in \design according to the requirements of HCNNG and TOGG. We also add another hardware baseline, CPU-T, which pairs the CPU with TeraByte-level DRAM, to see whether the optimized algorithms can benefit from larger main memory. From Fig.~\ref{HCNNG}, we can firstly observe that \design still outperforms other platforms even on these two more optimized algorithms. This is because the irregular and frequent data access still dominates the overhead of the search phases of HCNNG and TOGG though their searches are more directional. Secondly, we can see that although expanding the main memory can accelerate the search phase due to the lower data access latency, e.g., achieving 5.3$\times$ speedup over CPU with limited memory, CPU-T cannot beat in-storage accelerators because (\romannumeral1) DRAM fails to exploit the data locality and cannot match the high internal bandwidth of DeepStore and \design due to the lack of the in-memory logic
and  (\romannumeral2) CPU lacks the parallelism required to rival that of DeepStore or \design in search operations.


\subsection{Limitations of this work}
\label{limitation}
\design is proposed for the acceleration of graph-traversal-based ANNS algorithms but not generalized to some other types of ANNS algorithms like quantization-based ANNS~\cite{an2019quarter}~\cite{faiss} or tree-based ANNS~\cite{wang2013trinary}. We choose graph-traversal-based ANNS because it is currently the mainstream ANNS method~\cite{wang2021comprehensive}. In addition, \design also shows potential to be generalized to all the ANNS algorithms because all these ANNS workloads are memory-bound and their performance is limited by the memory bandwidth. \design can address these challenges fundamentally. We leave generalization of \design for future discussion.

\section{Related works}
\label{sec:relatedworks}
In-storage computing architectures aim to offload the data-intensive workloads to the storage devices like SSD to reduce the expensive data movement from and to storage. Modifying the internal architecture of SSD is required to develop an in-storage accelerator. Three categories of in-storage computing architectures have been proposed: \ding{182} modifying the firmware of the SSD controller (embedded cores)~\cite{wilkening2021recssd, graphssd}; \ding{183} adding customized hardware modules inside the SSD without touching the NAND flash chips~\cite{ glist, graphboost, mailthody2019deepstore, cognitivessd, inspire}; \ding{184} changing the internal architecture of NAND flash chips, e.g., adjusting the latching circuit or adding the digital logic in flash memory~\cite{gao2021parabit,gcd,ice}. 
Differing from the prior works like DeepStore~\cite{mailthody2019deepstore} which utilize the SSD internal bandwidth through simultaneous accesses to different flash channels, our work further improves the bandwidth of each flash channel by the modified multi-LUN operations and the two-level scheduling mechanism that exploit both the temporal and spatial data locality in all the page buffers, as shown in Fig.~\ref{Roofline}(b). Meanwhile, the data movement from flash channels is also reduced within the SSD.
GraphSSD~\cite{graphssd} considers the graph structure while deciding graph layout, access and update mechanisms mainly from the perspectives of the SSD controller and NVMe interface. 
GraphBoost~\cite{graphboost} only develops a sort-reduce accelerator between the Flash storage and DRAM. Both GraphSSD and GraphBoost do not conduct in-depth exploration of  the characteristics of the data access pattern in ANNS. Thus, they lacks the insight and design of 
exploiting the SSD internal bandwidth and page-level data locality. The SmartSSD-based solution~\cite{smartSSDKNN} connects an FPGA with an SSD via a PCIe switch and does not develop any in-storage logic inside the SSD. Hence, the performance of ~\cite{smartSSDKNN} is still limited by the low PCIe bandwidth. Implementing SmartSSD with the static scheduling in \design cannot fundamentally address the issue of PCIe bandwidth though the accesses to different pages in the SSD can be reduced.

\section{Conclusion}
\label{sec:conclusion}
In this paper, we present NDSearch, a novel NDP architecture based on SmartSSD, to support the graph-traversal-based ANNS task.
We present an in-storage accelerator \ssd which exploits the LUN-level parallelism inside NAND flash chips to utilize the high internal bandwidth of SSD.
We also  propose a customized processing model customized for NDP scenarios and implement it on our design to accelerate the graph-traversal-based ANNS.
Compared with the previous state-of-the-art NDP designs, \design achieves significant improvements in both throughput and energy efficiency.
\section{Acknowledgement}
\label{sec:acknowledgement}
This work was supported by NSF 1822085 and the membership from Samsung.

\bibliographystyle{IEEEtranS}
\bibliography{ref}

\begin{thebibliography}{10}
\providecommand{\url}[1]{#1}
\csname url@samestyle\endcsname
\providecommand{\newblock}{\relax}
\providecommand{\bibinfo}[2]{#2}
\providecommand{\BIBentrySTDinterwordspacing}{\spaceskip=0pt\relax}
\providecommand{\BIBentryALTinterwordstretchfactor}{4}
\providecommand{\BIBentryALTinterwordspacing}{\spaceskip=\fontdimen2\font plus
\BIBentryALTinterwordstretchfactor\fontdimen3\font minus \fontdimen4\font\relax}
\providecommand{\BIBforeignlanguage}[2]{{%
\expandafter\ifx\csname l@#1\endcsname\relax
\typeout{** WARNING: IEEEtranS.bst: No hyphenation pattern has been}%
\typeout{** loaded for the language `#1'. Using the pattern for}%
\typeout{** the default language instead.}%
\else
\language=\csname l@#1\endcsname
\fi
#2}}
\providecommand{\BIBdecl}{\relax}
\BIBdecl

\bibitem{bigann}
``{ANN} datasets,'' \url{https://big-ann-benchmarks.com}.

\bibitem{cacti}
``Cacti 6.5,'' \url{http://www.hpl.hp.com/research/cacti}.

\bibitem{cuhnsw}
``Cuhnsw,'' \url{https://github.com/js1010/cuhnsw}.

\bibitem{diskann}
``Disk{ANN},'' \url{https://github.com/microsoft/DiskANN}.

\bibitem{pm983}
``Download - simplessd 2.0,'' \url{https://docs.simplessd.org/en/v2.0.12/download.html#nvme-ssd}.

\bibitem{faiss}
``Faiss,'' \url{https://engineering.fb.com/2017/03/29/data-infrastructure/faiss-a-library-for-efficient-similarity-search/}.

\bibitem{GPUAcc}
``Gpuacc,'' \url{https://github.com/harsha-simhadri/big-ann-benchmarks/blob/main/t3/LEADERBOARDS_PUBLIC.md}.

\bibitem{hnswamazon}
``{HNSW} in {A}mazon,'' \url{https://aws.amazon.com/cn/blogs/database}.

\bibitem{hnswlib}
``Hnswlib - fast approximate nearest neighbor search,'' \url{https://github.com/nmslib/hnswlib}.

\bibitem{ssdsim}
``huaicheng/ssdsim,'' \url{https://github.com/huaicheng/ssdsim}.

\bibitem{vectordatabaseintro}
``Introduction of vector database,'' \url{https://www.pinecone.io/learn/vector-database/}.

\bibitem{LUNorgnization}
``Lun {O}rganization,'' \url{https://arxiv.org/pdf/1711.11427.pdf}.

\bibitem{milvus}
``Milvus,'' \url{https://milvus.io}.

\bibitem{rowaddress}
\BIBentryALTinterwordspacing
``Openssd {P}eer-to-{P}eer ({P2P}) - {X}ilinx {XRT}.'' [Online]. Available: \url{https://media-www.micron.com/-/media/client/onfi/specs/}
\BIBentrySTDinterwordspacing

\bibitem{rag}
``{R}etrieval {A}ugmented generation,'' \url{https://docs.aws.amazon.com/sagemaker/latest/dg/jumpstart-foundation-models-customize-rag.html}.

\bibitem{spacev}
``Spacev1b: A billion-scale vector dataset for text descriptors,'' \url{https://github.com/microsoft/SPTAG/tree/main/datasets/SPACEV1B}.

\bibitem{LUNorgnization2}
``{SSD} {S}pecification,'' \url{http://www.onfi.org/-/media/client/onfi/specs/onfi_4_1_gold.pdf?la=en}.

\bibitem{ssdsim2}
N.~Agrawal, V.~Prabhakaran, T.~Wobber, J.~D. Davis, M.~Manasse, and R.~Panigrahy, ``Design tradeoffs for ssd performance,'' in \emph{USENIX 2008 Annual Technical Conference}, ser. ATC'08.\hskip 1em plus 0.5em minus 0.4em\relax USA: USENIX Association, 2008, p. 57–70.

\bibitem{sift}
L.~Amsaleg and H.~Jégou, ``Datasets for approximate nearest neighbor search,'' \url{http://corpus-texmex.irisa.fr/}.

\bibitem{an2019quarter}
S.~An, Z.~Huang, S.~Bai, G.~Che, X.~Ma, J.~Luo, and Y.~Chen, ``Quarter-point product quantization for approximate nearest neighbor search,'' \emph{Pattern Recognition Letters}, vol. 125, pp. 187--194, 2019.

\bibitem{andoni2006near}
A.~Andoni and P.~Indyk, ``Near-optimal hashing algorithms for approximate nearest neighbor in high dimensions,'' in \emph{2006 47th annual IEEE symposium on foundations of computer science (FOCS'06)}.\hskip 1em plus 0.5em minus 0.4em\relax IEEE, 2006, pp. 459--468.

\bibitem{arya1998optimal}
S.~Arya, D.~M. Mount, N.~S. Netanyahu, R.~Silverman, and A.~Y. Wu, ``An optimal algorithm for approximate nearest neighbor searching fixed dimensions,'' \emph{Journal of the ACM (JACM)}, vol.~45, no.~6, pp. 891--923, 1998.

\bibitem{auroux2015reordering}
L.~Auroux, M.~Burelle, and R.~Erra, ``Reordering very large graphs for fun \& profit,'' in \emph{International Symposium on Web AlGorithms}, 2015.

\bibitem{babenko2016efficient}
A.~Babenko and V.~Lempitsky, ``Efficient indexing of billion-scale datasets of deep descriptors,'' in \emph{Proceedings of the IEEE Conference on Computer Vision and Pattern Recognition}, 2016, pp. 2055--2063.

\bibitem{barman2019graph}
A.~Barman and S.~K. Shah, ``A graph-based approach for making consensus-based decisions in image search and person re-identification,'' \emph{IEEE transactions on pattern analysis and machine intelligence}, 2019.

\bibitem{ml1}
Y.~Cao, H.~Qi, W.~Zhou, J.~Kato, K.~Li, X.~Liu, and J.~Gui, ``Binary hashing for approximate nearest neighbor search on big data: A survey,'' \emph{IEEE Access}, vol.~6, pp. 2039--2054, 2017.

\bibitem{caprara2005laying}
A.~Caprara and J.-J. Salazar-Gonz{\'a}lez, ``Laying out sparse graphs with provably minimum bandwidth,'' \emph{INFORMS Journal on Computing}, vol.~17, no.~3, pp. 356--373, 2005.

\bibitem{chen2018regan}
F.~Chen, L.~Song, and Y.~Chen, ``Regan: A pipelined reram-based accelerator for generative adversarial networks,'' in \emph{2018 23rd Asia and South Pacific Design Automation Conference (ASP-DAC)}.\hskip 1em plus 0.5em minus 0.4em\relax IEEE, 2018, pp. 178--183.

\bibitem{chen2015energy}
R.~Chen, S.~Siriyal, and V.~Prasanna, ``Energy and memory efficient mapping of bitonic sorting on fpga,'' in \emph{Proceedings of the 2015 ACM/SIGDA International Symposium on Field-Programmable Gate Arrays}, 2015, pp. 240--249.

\bibitem{chung2009survey}
T.-S. Chung, D.-J. Park, S.~Park, D.-H. Lee, S.-W. Lee, and H.-J. Song, ``A survey of flash translation layer,'' \emph{Journal of Systems Architecture}, vol.~55, no. 5-6, pp. 332--343, 2009.

\bibitem{coleman2021graph}
B.~Coleman, S.~Segarra, A.~Shrivastava, and A.~Smola, ``Graph reordering for cache-efficient near neighbor search,'' \emph{arXiv preprint arXiv:2104.03221}, 2021.

\bibitem{ml2}
S.~Cost and S.~Salzberg, ``A weighted nearest neighbor algorithm for learning with symbolic features,'' \emph{Machine learning}, vol.~10, pp. 57--78, 1993.

\bibitem{patternrecog1}
T.~Cover and P.~Hart, ``Nearest neighbor pattern classification,'' \emph{IEEE transactions on information theory}, vol.~13, no.~1, pp. 21--27, 1967.

\bibitem{eshghi2013ssd}
K.~Eshghi and R.~Micheloni, ``Ssd architecture and pci express interface,'' in \emph{Inside solid state drives (SSDs)}.\hskip 1em plus 0.5em minus 0.4em\relax Springer, 2013, pp. 19--45.

\bibitem{faultinjection}
B.~Fang, D.~Wang, S.~Jin, Q.~Koziol, Z.~Zhang, Q.~Guan, S.~Byna, S.~Krishnamoorthy, and D.~Tao, ``Characterizing impacts of storage faults on hpc applications: A methodology and insights,'' in \emph{2021 IEEE International Conference on Cluster Computing (CLUSTER)}.\hskip 1em plus 0.5em minus 0.4em\relax IEEE, 2021, pp. 409--420.

\bibitem{inforretr1}
M.~Flickner, H.~Sawhney, W.~Niblack, J.~Ashley, Q.~Huang, B.~Dom, M.~Gorkani, J.~Hafner, D.~Lee, D.~Petkovic \emph{et~al.}, ``Query by image and video content: The qbic system,'' \emph{computer}, vol.~28, no.~9, pp. 23--32, 1995.

\bibitem{fu2019fast}
C.~Fu, C.~Xiang, C.~Wang, and D.~Cai, ``Fast approximate nearest neighbor search with the navigating spreading-out graph,'' \emph{Proceedings of the VLDB Endowment}, vol.~12, no.~5, pp. 461--474, 2019.

\bibitem{fukunaga1975branch}
K.~Fukunaga and P.~M. Narendra, ``A branch and bound algorithm for computing k-nearest neighbors,'' \emph{IEEE transactions on computers}, vol. 100, no.~7, pp. 750--753, 1975.

\bibitem{gao2021parabit}
C.~Gao, X.~Xin, Y.~Lu, Y.~Zhang, J.~Yang, and J.~Shu, ``Parabit: Processing parallel bitwise operations in nand flash memory based ssds,'' in \emph{MICRO-54: 54th Annual IEEE/ACM International Symposium on Microarchitecture}, 2021, pp. 59--70.

\bibitem{csr}
A.~George, M.~T. Heath, J.~Liu, and E.~Ng, ``Solution of sparse positive definite systems on a shared-memory multiprocessor,'' \emph{International journal of parallel programming}, vol.~15, no.~4, pp. 309--325, 1986.

\bibitem{ice}
H.-W. Hu, W.-C. Wang, Y.-H. Chang, Y.-C. Lee, B.-R. Lin, H.-M. Wang, Y.-P. Lin, Y.-M. Huang, C.-Y. Lee, T.-H. Su, C.-C. Hsieh, C.-M. Hu, Y.-T. Lai, C.-K. Chen, H.-S. Chen, H.-P. Li, T.-W. Kuo, M.-F. Chang, K.-C. Wang, C.-H. Hung, and C.-Y. Lu, ``Ice: An intelligent cognition engine with 3d nand-based in-memory computing for vector similarity search acceleration,'' in \emph{2022 55th IEEE/ACM International Symposium on Microarchitecture (MICRO)}, 2022, pp. 763--783.

\bibitem{datamining1}
Q.~Huang, J.~Feng, Q.~Fang, W.~Ng, and W.~Wang, ``Query-aware locality-sensitive hashing scheme for lp norm,'' \emph{The VLDB Journal}, vol.~26, no.~5, pp. 683--708, 2017.

\bibitem{datamining2}
M.~Iwasaki, ``Pruned bi-directed k-nearest neighbor graph for proximity search,'' in \emph{Similarity Search and Applications: 9th International Conference, SISAP 2016, Tokyo, Japan, October 24-26, 2016, Proceedings 9}.\hskip 1em plus 0.5em minus 0.4em\relax Springer, 2016, pp. 20--33.

\bibitem{hard-decision}
R.~Jose and A.~Pe, ``Analysis of hard decision and soft decision decoding algorithms of ldpc codes in awgn,'' in \emph{2015 IEEE International Advance Computing Conference (IACC)}.\hskip 1em plus 0.5em minus 0.4em\relax IEEE, 2015, pp. 430--435.

\bibitem{graphboost}
\BIBentryALTinterwordspacing
S.-W. Jun, A.~Wright, S.~Zhang, S.~Xu, and Arvind, ``Grafboost: Using accelerated flash storage for external graph analytics,'' in \emph{Proceedings of the 45th Annual International Symposium on Computer Architecture}, ser. ISCA '18.\hskip 1em plus 0.5em minus 0.4em\relax IEEE Press, 2018, p. 411–424. [Online]. Available: \url{https://doi.org/10.1109/ISCA.2018.00042}
\BIBentrySTDinterwordspacing

\bibitem{llm}
E.~Kasneci, K.~Se{\ss}ler, S.~K{\"u}chemann, M.~Bannert, D.~Dementieva, F.~Fischer, U.~Gasser, G.~Groh, S.~G{\"u}nnemann, E.~H{\"u}llermeier \emph{et~al.}, ``Chatgpt for good? on opportunities and challenges of large language models for education,'' \emph{Learning and individual differences}, vol. 103, p. 102274, 2023.

\bibitem{smartSSDKNN}
J.-H. Kim, Y.-R. Park, J.~Do, S.-Y. Ji, and J.-Y. Kim, ``Accelerating large-scale graph-based nearest neighbor search on a computational storage platform,'' \emph{IEEE Transactions on Computers}, pp. 1--1, 2022.

\bibitem{soft-decision}
R.~Koetter and A.~Vardy, ``Algebraic soft-decision decoding of reed-solomon codes,'' \emph{IEEE Transactions on Information Theory}, vol.~49, no.~11, pp. 2809--2825, 2003.

\bibitem{patternrecog2}
A.~Kosuge and T.~Oshima, ``An object-pose estimation acceleration technique for picking robot applications by using graph-reusing k-nn search,'' in \emph{2019 First International Conference on Graph Computing (GC)}.\hskip 1em plus 0.5em minus 0.4em\relax IEEE, 2019, pp. 68--74.

\bibitem{lee2020smartssd}
J.~H. Lee, H.~Zhang, V.~Lagrange, P.~Krishnamoorthy, X.~Zhao, and Y.~S. Ki, ``Smartssd: Fpga accelerated near-storage data analytics on ssd,'' \emph{IEEE Computer architecture letters}, vol.~19, no.~2, pp. 110--113, 2020.

\bibitem{NP-C}
H.~R. Lewis, ``Michael r. $\pi$garey and david s. johnson. computers and intractability. a guide to the theory of np-completeness. wh freeman and company, san francisco1979, x+ 338 pp.'' \emph{The Journal of Symbolic Logic}, vol.~48, no.~2, pp. 498--500, 1983.

\bibitem{glist}
\BIBentryALTinterwordspacing
C.~Li, Y.~Wang, C.~Liu, S.~Liang, H.~Li, and X.~Li, ``{GLIST}: Towards {In-Storage} graph learning,'' in \emph{2021 USENIX Annual Technical Conference (USENIX ATC 21)}.\hskip 1em plus 0.5em minus 0.4em\relax USENIX Association, Jul. 2021, pp. 225--238. [Online]. Available: \url{https://www.usenix.org/conference/atc21/presentation/li-cangyuan}
\BIBentrySTDinterwordspacing

\bibitem{li2024ndrec}
S.~Li, Y.~Wang, E.~Hanson, A.~Chang, Y.~S. Ki, H.~H. Li, and Y.~Chen, ``Ndrec: A near-data processing system for training large-scale recommendation models,'' \emph{IEEE Transactions on Computers}, no.~01, pp. 1--14, 2024.

\bibitem{cognitivessd}
\BIBentryALTinterwordspacing
S.~Liang, Y.~Wang, Y.~Lu, Z.~Yang, H.~Li, and X.~Li, ``Cognitive {SSD}: A deep learning engine for {In-Storage} data retrieval,'' in \emph{2019 USENIX Annual Technical Conference (USENIX ATC 19)}.\hskip 1em plus 0.5em minus 0.4em\relax Renton, WA: USENIX Association, Jul. 2019, pp. 395--410. [Online]. Available: \url{https://www.usenix.org/conference/atc19/presentation/liang}
\BIBentrySTDinterwordspacing

\bibitem{inspire}
\BIBentryALTinterwordspacing
J.~Lin, L.~Liang, Z.~Qu, I.~Ahmad, L.~Liu, F.~Tu, T.~Gupta, Y.~Ding, and Y.~Xie, ``Inspire: In-storage private information retrieval via protocol and architecture co-design,'' in \emph{Proceedings of the 49th Annual International Symposium on Computer Architecture}, ser. ISCA '22.\hskip 1em plus 0.5em minus 0.4em\relax New York, NY, USA: Association for Computing Machinery, 2022, p. 102–115. [Online]. Available: \url{https://doi.org/10.1145/3470496.3527433}
\BIBentrySTDinterwordspacing

\bibitem{annsappdomain}
T.~Liu, A.~Moore, K.~Yang, and A.~Gray, ``An investigation of practical approximate nearest neighbor algorithms,'' \emph{Advances in neural information processing systems}, vol.~17, 2004.

\bibitem{two-stage-recsys}
J.~Ma, Z.~Zhao, X.~Yi, J.~Yang, M.~Chen, J.~Tang, L.~Hong, and E.~H. Chi, ``Off-policy learning in two-stage recommender systems,'' in \emph{Proceedings of The Web Conference 2020}, 2020, pp. 463--473.

\bibitem{mailthody2019deepstore}
V.~S. Mailthody, Z.~Qureshi, W.~Liang, Z.~Feng, S.~G. De~Gonzalo, Y.~Li, H.~Franke, J.~Xiong, J.~Huang, and W.-m. Hwu, ``Deepstore: In-storage acceleration for intelligent queries,'' in \emph{Proceedings of the 52nd Annual IEEE/ACM International Symposium on Microarchitecture}, 2019, pp. 224--238.

\bibitem{malkov2018efficient}
Y.~A. Malkov and D.~A. Yashunin, ``Efficient and robust approximate nearest neighbor search using hierarchical navigable small world graphs,'' \emph{IEEE transactions on pattern analysis and machine intelligence}, vol.~42, no.~4, pp. 824--836, 2018.

\bibitem{graphssd}
\BIBentryALTinterwordspacing
K.~K. Matam, G.~Koo, H.~Zha, H.-W. Tseng, and M.~Annavaram, ``Graphssd: Graph semantics aware ssd,'' in \emph{Proceedings of the 46th International Symposium on Computer Architecture}, ser. ISCA '19.\hskip 1em plus 0.5em minus 0.4em\relax New York, NY, USA: Association for Computing Machinery, 2019, p. 116–128. [Online]. Available: \url{https://doi.org/10.1145/3307650.3322275}
\BIBentrySTDinterwordspacing

\bibitem{cvpr20_tutorial_image_retrieval}
Y.~Matsui, T.~Yamaguchi, and Z.~Wang, ``Cvpr2020 tutorial on image retrieval in the wild,'' \url{https://matsui528.github.io/cvpr2020_tutorial_retrieval/}, 2020.

\bibitem{recsys1}
Y.~Meng, X.~Dai, X.~Yan, J.~Cheng, W.~Liu, J.~Guo, B.~Liao, and G.~Chen, ``Pmd: An optimal transportation-based user distance for recommender systems,'' in \emph{Advances in Information Retrieval: 42nd European Conference on IR Research, ECIR 2020, Lisbon, Portugal, April 14--17, 2020, Proceedings, Part II 42}.\hskip 1em plus 0.5em minus 0.4em\relax Springer, 2020, pp. 272--280.

\bibitem{hcnng}
J.~V. Munoz, M.~A. Gon{\c{c}}alves, Z.~Dias, and R.~d.~S. Torres, ``Hierarchical clustering-based graphs for large scale approximate nearest neighbor search,'' \emph{Pattern Recognition}, vol.~96, p. 106970, 2019.

\bibitem{gcd}
W.~Niu, J.~Guan, X.~Shen, Y.~Wang, G.~Agrawal, and B.~Ren, ``Gcd2: A globally optimizing compiler for mapping dnns to mobile dsps,'' in \emph{2022 55th IEEE/ACM International Symposium on Microarchitecture (MICRO)}, 2022, pp. 512--529.

\bibitem{pennington2014glove}
J.~Pennington, R.~Socher, and C.~D. Manning, ``Glove: Global vectors for word representation,'' in \emph{Proceedings of the 2014 conference on empirical methods in natural language processing (EMNLP)}, 2014, pp. 1532--1543.

\bibitem{salamat2021nascent}
S.~Salamat, A.~Haj~Aboutalebi, B.~Khaleghi, J.~H. Lee, Y.~S. Ki, and T.~Rosing, ``Nascent: Near-storage acceleration of database sort on smartssd,'' in \emph{The 2021 ACM/SIGDA International Symposium on Field-Programmable Gate Arrays}, 2021, pp. 262--272.

\bibitem{recsys2}
B.~Sarwar, G.~Karypis, J.~Konstan, and J.~Riedl, ``Item-based collaborative filtering recommendation algorithms,'' in \emph{Proceedings of the 10th international conference on World Wide Web}, 2001, pp. 285--295.

\bibitem{semiconductor2006open}
H.~Semiconductor \emph{et~al.}, ``Open nand flash interface specification,'' \emph{Technical Report ONFI}, 2006.

\bibitem{stanley1986enumerative}
R.~P. Stanley, ``Enumerative combinatorics, volume 1. wadsworth,'' \emph{Inc. California}, 1986.

\bibitem{subramanya2019diskann}
S.~J. Subramanya, F.~Devvrit, H.~Simhadri, R.~Krishnawamy, and R.~Kadekodi, ``Diskann: Fast accurate billion-point nearest neighbor search on a single node,'' \emph{Advances in Neural Information Processing Systems}, vol.~32, pp. 13\,771--13\,781, 2019.

\bibitem{sundaram2015graphmat}
N.~Sundaram, N.~Satish, M.~M.~A. Patwary, S.~R. Dulloor, M.~J. Anderson, S.~G. Vadlamudi, D.~Das, and P.~Dubey, ``Graphmat: high performance graph analytics made productive,'' \emph{Proceedings of the VLDB Endowment}, vol.~8, no.~11, pp. 1214--1225, 2015.

\bibitem{ldpc}
T.~Tian, C.~R. Jones, J.~D. Villasenor, and R.~D. Wesel, ``Selective avoidance of cycles in irregular ldpc code construction,'' \emph{IEEE Transactions on Communications}, vol.~52, no.~8, pp. 1242--1247, 2004.

\bibitem{wang2013trinary}
J.~Wang, N.~Wang, Y.~Jia, J.~Li, G.~Zeng, H.~Zha, and X.-S. Hua, ``Trinary-projection trees for approximate nearest neighbor search,'' \emph{IEEE transactions on pattern analysis and machine intelligence}, vol.~36, no.~2, pp. 388--403, 2013.

\bibitem{wang2021comprehensive}
M.~Wang, X.~Xu, Q.~Yue, and Y.~Wang, ``A comprehensive survey and experimental comparison of graph-based approximate nearest neighbor search,'' \emph{arXiv preprint arXiv:2101.12631}, 2021.

\bibitem{wang2020reboc}
Y.~Wang, F.~Chen, L.~Song, C.-J.~R. Shi, H.~H. Li, and Y.~Chen, ``Reboc: Accelerating block-circulant neural networks in reram,'' in \emph{2020 Design, Automation \& Test in Europe Conference \& Exhibition (DATE)}.\hskip 1em plus 0.5em minus 0.4em\relax IEEE, 2020, pp. 1472--1477.

\bibitem{wang2023ems}
Y.~Wang, S.~Li, Q.~Zheng, A.~Chang, H.~Li, and Y.~Chen, ``Ems-i: An efficient memory system design with specialized caching mechanism for recommendation inference,'' \emph{ACM Transactions on Embedded Computing Systems}, vol.~22, no.~5s, pp. 1--22, 2023.

\bibitem{wang2021rerec}
Y.~Wang, Z.~Zhu, F.~Chen, M.~Ma, G.~Dai, Y.~Wang, H.~Li, and Y.~Chen, ``Rerec: In-reram acceleration with access-aware mapping for personalized recommendation,'' in \emph{2021 IEEE/ACM International Conference On Computer Aided Design (ICCAD)}.\hskip 1em plus 0.5em minus 0.4em\relax IEEE, 2021, pp. 1--9.

\bibitem{wen2019memristor}
S.~Wen, H.~Wei, Z.~Yan, Z.~Guo, Y.~Yang, T.~Huang, and Y.~Chen, ``Memristor-based design of sparse compact convolutional neural network,'' \emph{IEEE Transactions on Network Science and Engineering}, vol.~7, no.~3, pp. 1431--1440, 2019.

\bibitem{wilkening2021recssd}
M.~Wilkening, U.~Gupta, S.~Hsia, C.~Trippel, C.-J. Wu, D.~Brooks, and G.-Y. Wei, ``Recssd: near data processing for solid state drive based recommendation inference,'' in \emph{Proceedings of the 26th ACM International Conference on Architectural Support for Programming Languages and Operating Systems}, 2021, pp. 717--729.

\bibitem{xiao2017fashion}
H.~Xiao, K.~Rasul, and R.~Vollgraf, ``Fashion-mnist: a novel image dataset for benchmarking machine learning algorithms,'' \emph{arXiv preprint arXiv:1708.07747}, 2017.

\bibitem{togg}
X.~Xu, M.~Wang, Y.~Wang, and D.~Ma, ``Two-stage routing with optimized guided search and greedy algorithm on proximity graph,'' \emph{Knowledge-Based Systems}, vol. 229, p. 107305, 2021.

\bibitem{yan2019alleviating}
M.~Yan, X.~Hu, S.~Li, A.~Basak, H.~Li, X.~Ma, I.~Akgun, Y.~Feng, P.~Gu, L.~Deng \emph{et~al.}, ``Alleviating irregularity in graph analytics acceleration: a hardware/software co-design approach,'' in \emph{Proceedings of the 52nd Annual IEEE/ACM International Symposium on Microarchitecture}, 2019, pp. 615--628.

\bibitem{ber}
K.~Zhao, W.~Zhao, H.~Sun, X.~Zhang, N.~Zheng, and T.~Zhang, ``{LDPC-in-SSD}: Making advanced error correction codes work effectively in solid state drives,'' in \emph{11th USENIX Conference on File and Storage Technologies (FAST 13)}, 2013, pp. 243--256.

\bibitem{two-stage-reid}
Z.~Zheng, X.~Yang, Z.~Yu, L.~Zheng, Y.~Yang, and J.~Kautz, ``Joint discriminative and generative learning for person re-identification,'' in \emph{proceedings of the IEEE/CVF conference on computer vision and pattern recognition}, 2019, pp. 2138--2147.

\bibitem{inforetr2}
C.~J. Zhu, T.~Zhu, H.~Li, J.~Bi, and M.~Song, ``Accelerating large-scale molecular similarity search through exploiting high performance computing,'' in \emph{2019 IEEE International Conference on Bioinformatics and Biomedicine (BIBM)}.\hskip 1em plus 0.5em minus 0.4em\relax IEEE, 2019, pp. 330--333.

\bibitem{zhu2016deep}
H.~Zhu, M.~Long, J.~Wang, and Y.~Cao, ``Deep hashing network for efficient similarity retrieval,'' in \emph{Proceedings of the AAAI Conference on Artificial Intelligence}, vol.~30, no.~1, 2016.

\end{thebibliography}

\end{document}